\documentclass[12pt,preprint]{aastex}
\usepackage{amsmath}
\usepackage{amsfonts}
\usepackage{amssymb}
\usepackage{float}
\usepackage{graphicx}
\usepackage{calc}
\usepackage{listings}
\usepackage{hyperref}
\usepackage{color, soul}

\usepackage{color}

\newcommand\grim{{\tt grim}}

\def\bh{\hat{b}}

\begin{document}

\title{\grim: A Flexible, Conservative Scheme for Relativistic Fluid Theories}

\author{Mani Chandra}
\affil{Department of Astronomy, University of Illinois, 1110 West Green Street, Urbana, IL, 61801; {\tt manic@illinois.edu}}

\author{Francois Foucart}
\affil{Lawrence Berkeley National Laboratory, 1 Cyclotron Rd, Berkeley, CA 94720, USA; Einstein Fellow; {\tt fvfoucart@lbl.gov}}

\author{Charles F. Gammie}
\affil{Department of Astronomy, University of Illinois, 1002 West Green Street, Urbana, IL, 61801}
\affil{Department of Physics, University of Illinois, 1110 West Green Street, Urbana, IL, 61801; {\tt gammie@illinois.edu}}

\begin{abstract}

Hot, diffuse, relativistic plasmas such as sub-Eddington black hole accretion flows are expected to be collisionless, yet are commonly modeled as a fluid using ideal general relativistic magnetohydrodynamics (GRMHD). Dissipative effects such as heat conduction and viscosity can be important in a collisionless plasma and will potentially alter the dynamics and radiative properties of the flow from that in ideal fluid models; we refer to models that include these processes as Extended GRMHD. Here we describe a new conservative code, \grim\footnote{General Relativistic \emph{Implicit} Magnetohydrodynamics: \url{http://github.com/afd-illinois/grim}. Commit hash used in this paper: {\tt 70bcd77}}, that enables all the above and additional physics to be efficiently incorporated. \grim~combines time evolution and primitive variable inversion needed for conservative schemes into a single step using an algorithm that only requires the residuals of the governing equations as inputs. This algorithm enables the code to be physics agnostic as well as flexibility regarding time-stepping schemes. \grim~runs on CPUs, as well as on GPUs, using the \emph{same} code. We formulate a performance model, and use it to show that our implementation runs optimally on both architectures. \grim~correctly captures classical GRMHD test problems as well as a new suite of linear and nonlinear test problems with anisotropic conduction and viscosity in special and general relativity.  As tests and example applications, we resolve the shock substructure due to the presence of dissipation, and report on relativistic versions of the magneto-thermal instability and heat flux driven buoyancy instability, which arise due to anisotropic heat conduction, and of the firehose instability, which occurs due to anisotropic pressure (i.e. viscosity). Finally, we show an example integration of an accretion flow around a Kerr black hole, using Extended GRMHD.

\end{abstract}

\section{Introduction}

The fluid description of a plasma using the ideal general relativistic
magnetohydrodynamic (GRMHD) equations is a workhorse in theoretical high energy
astrophysics. The codes that solve these equations have been successfully
applied in studies of various processes of interest such as jet formation and
accretion onto compact objects. Many important results have emerged from
numerical solutions of the ideal GRMHD equations. A few examples are the
validation that the \citealt{BZ} mechanism occurs naturally in a global MHD
model \citep{mckinney2004}, the discovery of magnetically chocked accretion
flows \citep{mckinney2012, MADJet}, and simulated observations of Sgr A*
\citep{monika2009}.

However, the ideal GRMHD model is readily justified only when the Knudsen number $Kn = l_{mfp}/l_{system} \ll 1$, where $l_{mfp}$ is the mean free path, and $l_{system}$ is the characteristic length scale of the system, and when the ratio of the time scales $\tau_{C}/\tau_{D} \ll 1$, where $\tau_C$ is the two-body Coulomb scattering time scale, and $\tau_D$ is the dynamical time scale in the system. In other words, the ideal GRMHD model assumes that the plasma is locally in equilibrium. This leads to a simple set of conservation laws for mass and momentum and all that is required to complete the system is a prescription for the pressure, which is usually approximated by a Gamma-law equation of state. While this simplicity is appealing, systems such as low luminosity black holes which accrete through a radiatively inefficient accretion flow (RIAF) are in the $Kn \gg 1$ regime.

In a RIAF, the synchrotron cooling time scales are much longer than the dynamical time scale. This leads to the accreting plasma becoming virially hot as the gravitational potential energy is stored as internal energy, with $T \sim R^{-1}$, where $T$ is the temperature of the plasma, and $R$ is the radius from the black hole. The disk is then geometrically thick, and optically thin (\cite{Yuan2014}) and the Coulomb mean free paths between all the constituent particles (ion-ion, ion-electron, electron-electron) (all of which scale as $\sim T^2$) are much larger than the typical system scale $GM/c^2$ \citep{Mahadevan1997}. Thus, it is not evident that ideal GRMHD is applicable.

Despite the divergence of the Knudsen number, and the collisional time scale, there are indeed small parameters that can be exploited to recover an effective hydrodynamic description. In the presence of a sufficiently strong magnetic field, the following conditions can apply: $l_{gyro}/l_{system} \ll 1$, and $t_{gyro}/t_{system} \ll 1$, where $l_{gyro}$ is the gyroradius and $t_{gyro}$ is the gyroperiod. These apply in most astrophysical systems. For example, in Sgr A*, Faraday rotation measurements and observed synchrotron radiation indicate a magnetic field strength $\sim 100$ Gauss and number density $\sim 10^{7}$ $cm^{-3}$. implying $l_{\rm gyro}/l_{\rm system} \sim 10^{-5}$ and $t_{\rm gyro}/t_{\rm system} \sim 10^{-8}$. Thus, particles are constrained to move along field lines.  In the presence of weak collisionality, perhaps provided by wave-particle scattering, this leads to set of fluid-like equations with anisotropic transport along the magnetic field lines.

Dissipative relativistic fluid theories should be hyperbolic, causal, and
stable. Early theories by \citealt{Eckart1940} and Landau-Lifshitz do not
satisfy these requirements whereas these are conditionally satisfied by the
\citealt{Israel1979} theory of dissipative hydrodynamics \citep{Hiscock1983,
Hiscock1985, Hiscock1988, Hiscock1988b}.
\citealt{chandra2015} adapted the \citealt{Israel1979} theory for isotropic
conduction and viscosity, taking into account the symmetries imposed on the
distribution function of a plasma in the presence of a magnetic field to derive
a one-fluid model of a plasma that incorporates anisotropic thermal conduction
and viscosity. The conduction is driven by temperature gradients along field
lines and the viscosity due to a shear flow projected onto the field lines. The
model, referred to as extended magnetohydrodynamics (EMHD), is valid up to
second order deviations from equilibrium and is applicable to weakly collisional
flows. We review the equations of the model in section (\S
\ref{section:review_EMHD}) and encourage the interested reader to look at
\citealt{chandra2015} for the derivation and the limits of the model within
which it satisfies the above mentioned constraints. In this paper, we derive a
variety of analytic and semi-analytic solutions, described in (\S \ref{section:tests}), to develop intuition about the EMHD model, and to serve in a test suite for the numerical implementation of EMHD and similar models.

The methods used to integrate the equations of relativistic MHD are similar to
those used in non-relativistic MHD, namely, shock capturing conservative schemes
using the finite volume method. In particular, the approximate Riemann solvers
used to compute the numerical fluxes at cell interfaces, and the various methods
available to evolve the magnetic field under the constraint $\nabla\cdot
\mathbf{B} = 0$ are similar for relativistic and non-relativistic MHD. One of
the main complication in relativistic MHD is the mathematical relation between
the evolved variables and physical variables. Consider special-relativistic
ideal hydrodynamics, where the physical variables to be solved for, referred to
as \emph{primitive variables}, are the rest mass energy density $\rho$, the
internal energy $u$ and the spatial components of the four-velocity $u^i$. The
variables are evolved using the continuity equation $\partial_\mu(\rho
u^\mu)=0$, and the energy and momentum conservation equations given by
$\partial_\mu T^{\mu \nu} = 0$, where $T^{\mu \nu} = (\rho + u + P)u^\mu u^\nu +
P \eta^{\mu \nu}$ is the perfect fluid stress tensor, $m$ is the particle mass,
$\eta^{\mu\nu}=\rm{diag}(-1,1,1,1)$ is the flat space metric, and $P$ is the
pressure, approximated here by a gamma-law equation of state, $P_g = (\gamma - 1)u$. Conservative schemes time-step the \emph{conserved variables}, ${\bf U}=(\rho u^0,T^{0 \nu})$, from ${\bf U}^n$ to ${\bf U}^{n+1}$, where the superscripts $n$, $n+1$ indicate the discretized time levels. To recover the primitive variables $n^{n+1}$, $u^{n+1}$ and $(u^i)^{n+1}$ at the new time step from ${\bf U}^{n+1}$ requires the solution to a set of nonlinear equations and is a multivariate nonlinear root finding problem (although for hydrodynamics it can be reduced to a univariate nonlinear problem). This is unlike non-relativistic fluid dynamics, where this recovery step is algebraic.

Many schemes have been proposed for the recovery of primitive variables from conserved variables in relativistic hydrodynamics \citep{primToCons}. However, the introduction of new physics, as in the EMHD model, voids the earlier algorithms, which are specialized to ideal MHD . The model has equations governing the dissipative quantities $q$, the heat flux along the magnetic field lines, and $\Delta P$, the pressure anisotropy, which are of the form $\partial_t q \sim \hat{b}^\mu\partial_\mu T+ \hat{b}^\mu u^\nu\partial_\nu u_\mu$ and $\partial_t \Delta P \sim \hat{b}^\mu\hat{b}^\nu \partial_\mu u_\nu$, where $T$ is the temperature and $\hat{b}^\mu$ is the unit vector along the direction of the magnetic field. The difficulty is that the equations for $q$ and $\Delta P$ are sourced by \emph{spatio-temporal} derivatives and not just spatial derivatives. The values of $q^{n+1}$ and $\Delta P^{n+1}$ depend on the values of $T^{n+1}$ and $u^{n+1}_\nu$, but these in turn need to be recovered from the conserved quantities ${\bf U}^{n+1}$. Now, the stress-tensor has dissipative contributions of the form $T^{\mu \nu} \sim q (\hat{b}^\mu u^\nu + \hat{b}^\nu u^\mu) + \Delta P \hat{b}^\mu\hat{b}^\nu$ and so ${\bf U}^{n+1}$ itself depends on $q^{n+1}$ and $\Delta P^{n+1}$. Thus, the time evolution of all thermodynamic quantities are nonlinearly intertwined with primitive variable recovery.

{\tt grim} recasts the entire time stepping procedure as a coupled multivariate nonlinear root finding problem.  Consider as a simple example the following system of 1D wave-equations:
\begin{align}
\partial_t u_1 + c\partial_x u_1 & = 0 \\
\partial_t u_2 + c\partial_x u_2 & = 0
\end{align}
for the variables $u_1(x,t)$ and $u_2(x,t)$. Now, performing an explicit first order spatio-temporal discretization, we have $(u^{i,n+1}_{1,2} - u^{i,n}_{1,2})/\Delta t + c(u^{i+1,n}_{1,2} - u^{i,n}_{1,2})/\Delta x = 0$ (assuming $c>0$), where the index $i$ denotes a grid zone and the index $n$ denotes a time level. Here, both $u^{i,n+1}_1$ and $u^{i,n+1}_2$ can be solved for algebraically, $u^{i,n+1}_{1,2} = u^{i,n}_{1,2} - c\Delta t/\Delta x (u^{i+1,n}_{1,2} - u^{i,n}_{1,2})$. In \grim, we find instead the values of $u^{i,n+1}_1$ and $u^{i,n+1}_2$ that satisfy
\begin{equation}
        \mathbf{f}(u^{i,n+1}_1, u^{i,n+1}_2) \equiv \left\{
    \begin{aligned}
        (u^{i, n+1}_{1}  - u^{i,n}_{1})/\Delta t + c(u^{i+1,n}_{1} - u^{i,n}_{1})/\Delta x \\
        (u^{i, n+1}_{2}  - u^{i,n}_{2})/\Delta t + c(u^{i+1,n}_{2} - u^{i,n}_{2})/\Delta x
    \end{aligned}
\right\} = 0,
\end{equation}
where  $\mathbf{f}(u^{i,n+1}_1, u^{i,n+1}_2)$ are the \emph{residuals}, and represent the governing equations in their discretized form. This system of equations (which in general are nonlinearly coupled) is now solved using an iterative algorithm until $|\mathbf{f}(u^{i,n+1}_1, u^{i, n+1}_2)| < \epsilon$, where $|.|$ is a suitable norm and $\epsilon$ is a chosen tolerance. The algorithm requires as sole input the residuals $\mathbf{f}(...)$, which are the discretized form of the governing equations. The algorithm is independent of the physics that constitutes the discretized equations $\mathbf{f}(...)$ and is therefore independent of the underlying physical model. It works with ideal MHD, EMHD, and possible extensions of the EMHD model. Thus, the abstraction of numerical solution to a set of PDEs as a nonlinear root finding problem allows for flexibility regarding the governing equations, as well as time-stepping schemes as we shall show in later sections.

We begin in \S \ref{section:finite_volume_method} by describing the numerical
discretization of a set of hyperbolic PDEs to $O(\Delta x^2, \Delta t^2)$ using
the finite volume method combined with a semi-implicit time stepping scheme. We
then proceed in \S \ref{section:root_finder} to recast the time stepping of the
discrete system as a non-linear multivariate root finding problem and describe
how the roots are obtained using a residual-based algorithm. We then apply this
technique to the EMHD model in \S \ref{section:review_EMHD}, along with a review
of the governing equations. We detail the implementation of all the above in \S
\ref{section:implementation}, and describe the techniques, and libraries we use
that enable us to use either CPUs, or GPUs. We then report various performance
and scaling data in \S \ref{section:scaling}. In order to understand the
performance numbers, we formulate a performance model in \S
\ref{section:performance}, and use it to show that our implementation is optimal
on both CPUs, and GPUs. We have developed an extensive test suite for the EMHD
model which we present in \S \ref{section:tests}, and validate {\tt grim} using
this test suite, thus demonstrating its utility in exploring the solution space
of this model. In \S \ref{section:applications}, we show example applications of
{\tt grim}; buoyancy instabilities that occur in weakly collisional plasmas, and
accretion onto supermassive black holes. Finally, in \S
\S \ref{section:conclusion}, we conclude.

\section{Finite volume method} \label{section:finite_volume_method}

{\tt grim} uses the finite volume method to solve hyperbolic partial differential equations in their conservative form
\begin{align}
\partial_t U + \partial_j F^j & = S \label{eq:conservative_eqns}
\end{align}
where $U$ is the vector of \emph{conserved} quantities, $F^j$ are \emph{fluxes}, and $S$ are \emph{sources}. We break down the full scheme into (\S \ref{section:grid_generation}) domain discretization, (\S \ref{section:integral_form}) integral form of the differential equations, (\S \ref{section:time_stepping}) time stepping scheme, and (\S \ref{section:spatial_discretization}) spatial discretization.

\subsection{Grid Generation} \label{section:grid_generation}

We are primarily interested in solving (\ref{eq:conservative_eqns}) in simple rectangular and spherical geometries. To discretize these domains, we work in coordinates where the boundaries of the domains are aligned with the coordinate axes. For example, Cartesian coordinates $x^i = \{x, y, z\}$ for rectangular domains, and spherical polar coordinates $x^i = \{r, \theta, \phi\}$ for spherical geometries. Then, given the extent of the domain in these coordinates $\left[x^i_{start}, x^i_{end}\right]$, a grid with $N_1 \times N_2 \times N_3$ zones is generated by decomposing the spatial domain into zones with dimensions $dx^1 \times dx^2 \times dx^3$, where $dx^i =(x^i_{end} - x^i_{start})/N^i$, for $i=1, 2, 3$. This results in a uniform mesh in each coordinate.

If a physical problem requires concentration of grid zones in a specific region, we construct a smooth curvilinear non-uniform grid using a coordinate transformation, as is done in the {\tt harm} code \citep{harm}. First, a uniform grid is generated in a different set of coordinates $X^i$, and then transformed to the $x^i$ coordinates using $x^i \equiv x^i(X^j)$. The grid zones in $X^i$ all have equal dimensions $dX^1 \times dX^2 \times dX^3$, where $dX^i =(X^i_{end} - X^i_{start})/N^i$, and $\left[X^i_{start}, X^i_{end}\right]$ is the extent of the domain in the new coordinates. This corresponds to a grid spacing $dx^i = L^i_j dX^j$ in $x^i$, where $L^i_j \equiv \partial x^i/\partial X^j$ is the transformation matrix. Depending on the form of $x^i(X^j)$, a non-uniform grid is generated in the $x^i$ coordinates.

Below, we illustrate the grid generation for a domain enclosed by two spherical shells. We concentrate the grid zones near the inner radius $r_{in}$ using a $\log(r)$ grid, and near the midplane $\theta = \pi/2$, with an adjustable parameter $h = (0, 1]$. As $h \rightarrow 0$, there is greater concentration of the zones near the midplane.
\begin{align}
x^1 \equiv r         & = \exp(X^1) \\
x^2 \equiv \theta    & = \pi X^2 + \left(\frac{1 - h}{2}\right)\sin(2\pi X^2) \label{eqn:theta_transform} \\
x^3 \equiv \phi      & = X^3 \\
L^i_j & =
\left(
\begin{array}{ccc}
\exp(X^1) & 0            & 0 \\
0         & \pi(1 + (1-h)\cos(2\pi X^2))       & 0 \\
0         & 0            & 1
\end{array}
\right)
\end{align}
The boundaries of the domain in $x^i = \{r, \theta, \phi\}$ are $[r_{in}, r_{out}] \times [0, \pi] \times [0, 2\pi]$, which correspond to $[\log(r_{in}), \log(r_{out})] \times [0, 1] \times [0, 1]$ in $X^i = \{X^1, X^2, X^3\}$ coordinates. 

\begin{figure}[!htbp]
\begin{center}
\includegraphics[width=150mm]{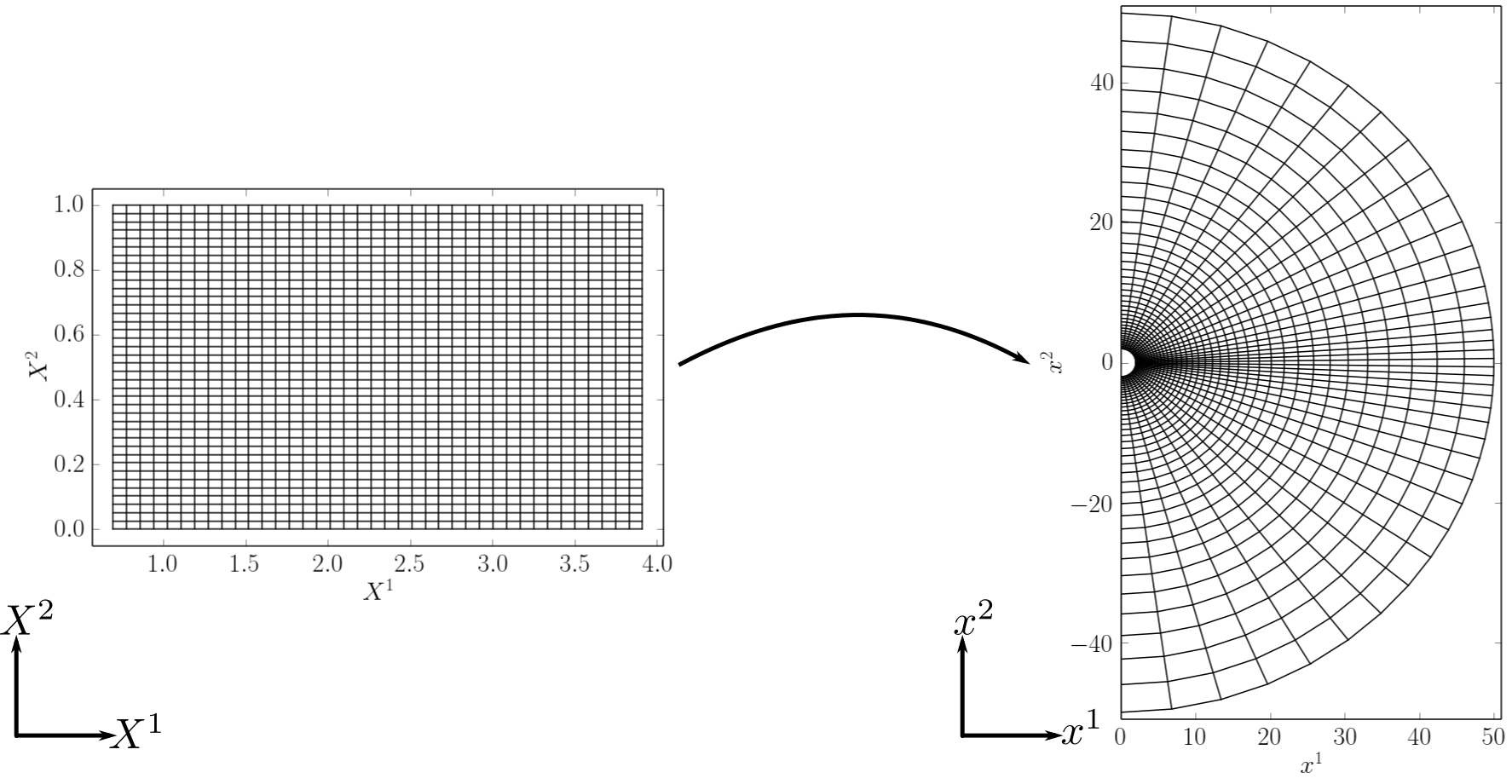}
\end{center}
\caption{Generation of a spherical grid with mid-plane grid refinement, with the
  refinement parameter $h = 0.3$ in (\ref{eqn:theta_transform}). The left side
  shows the grid in the computational coordinates $(X^1, X^2)$ that {\tt grim}
  works in, and the right side shows the grid in Cartesian coordinates.}
\label{fig:spherical_grid}
\end{figure}

\subsection{Integral Form of the Differential Equations} \label{section:integral_form}

\begin{figure}[!htbp]\label{fig:gridzone_schematic}
\begin{center}
\includegraphics[width=100mm]{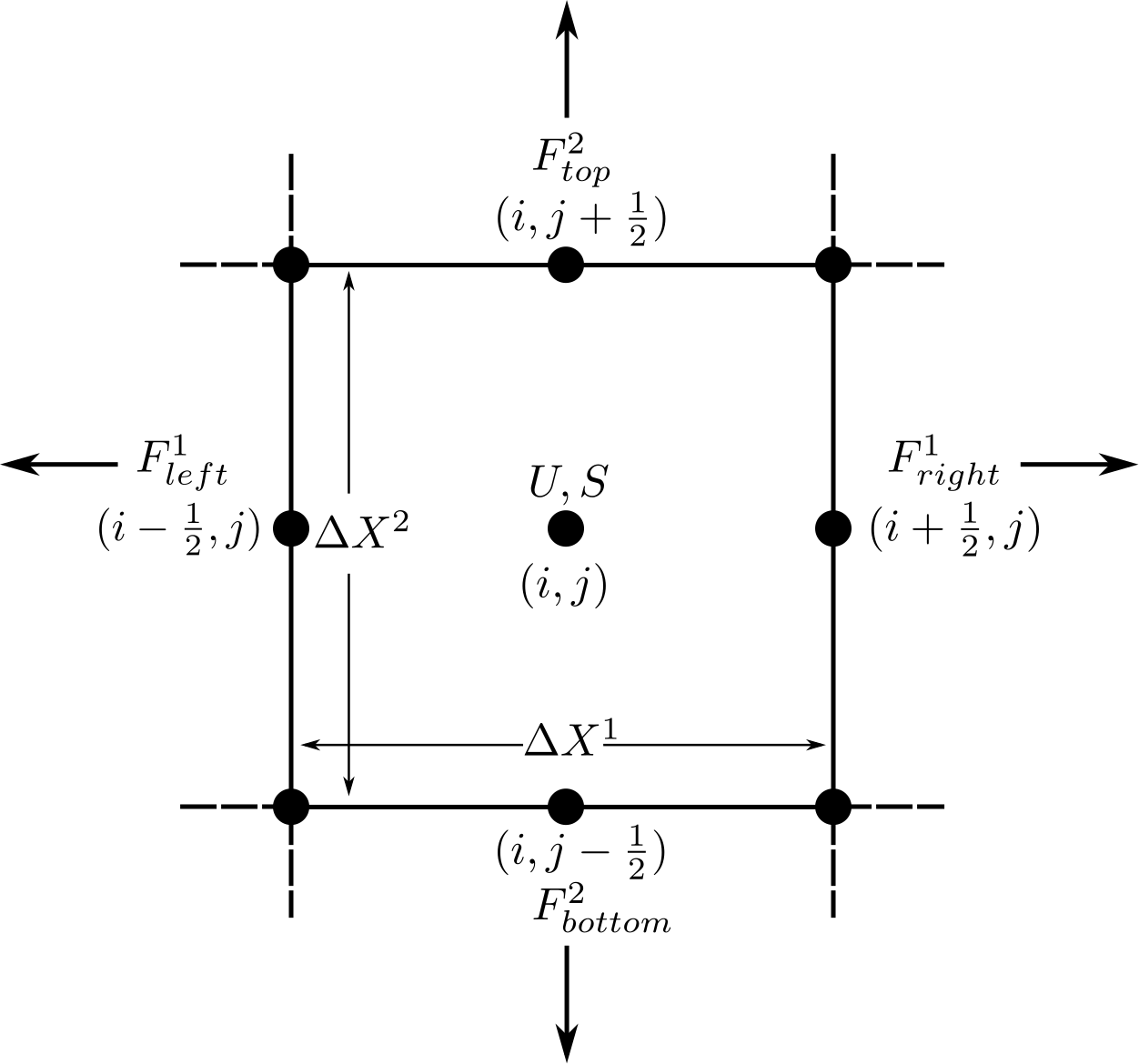}
\end{center}
\caption{Schematic of a grid zone in 2D. In 3D, there are two additional faces
  along $X3$, which we denote by {\it front} and {\it back}.}
\label{linear modes}
\end{figure}

We now setup the finite volume formulation in the $(X^1, X^2, X^3)$ coordinate system. 
Multiplying (\ref{eq:conservative_eqns}) by the area of the \emph{control volume} (see fig) $\Delta v = dX^1 dX^2 dX^3$, we get
\begin{align} \label{eq:fvm_integral_formulation}
\partial_t \int U \Delta v
+
\int dX^1\partial_1 \left(\int F^1 dX^2 dX^3\right) + \int dX^2\partial_2 \left(\int F^2 dX^1dX^3\right) + ...
 = \int S \Delta v
\end{align}
Rewriting the above in terms of \emph{cell-averages} $\bar{U} \equiv \int U
\Delta v/\int \Delta v$, $\bar{S} \equiv \int S \Delta v/\int \Delta v$ and the
\emph{face-averages} $\bar{F}^1 \equiv \int F^1 dX^2 dX^3/\int dX^2 dX^3$,
$\bar{F}^2 \equiv \int F^2 dX^1 dX^3/\int dX^1 dX^3$, and $\bar{F}^3 \equiv \int F^3
dX^1 dX^2/\int dX^1 dX^2$
\begin{align} \label{eq:fvm_formulation}
\partial_t \bar{U} + \frac{\bar{F}^1_{\rm right} - \bar{F}^1_{\rm left}}{\Delta X^1} + \frac{\bar{F}^2_{\rm top} - \bar{F}^2_{\rm bottom}}{\Delta X^2} + \frac{\bar{F}^3_{\rm front} - \bar{F}^3_{\rm back}}{\Delta X^3}= \bar{S}
\end{align}
where we have replaced $\int dX^1 \partial_1()$ in
(\ref{eq:fvm_integral_formulation}) by the surface integral of $F^1$ on the
$right$ and $left$ surfaces of the control volume (see
fig.~\ref{fig:gridzone_schematic}),  and $\int dX^2 \partial_2()$, $\int dX^3
\partial_3()$ have been replaced by surface integrals of $F^2$ on the $top$ and
$bottom$ surfaces, and similarly for $F^3$ on the $front$ and $back$ surfaces respectively. The above equations (\ref{eq:fvm_formulation}) are an \emph{exact} integral reformulation of the differential equations (\ref{eq:conservative_eqns}) over the control volume. Multiplying (\ref{eq:fvm_formulation}) by $\int dt$ and performing the integration over a discrete time interval $\Delta t$,
\begin{align} \label{eq:fvm_formulation_integral_dt}
\bar{U}_{n+1} - \bar{U}_n + \frac{\int dt \bar{F}^1_{\rm right} - \int dt\bar{F}^1_{\rm left}}{\Delta X^1} + \frac{\int dt\bar{F}^2_{\rm top} - \int dt\bar{F}^2_{\rm bottom}}{\Delta X^2} + ... = \int dt \bar{S}
\end{align}
where the index $n$ indicates the discrete time level. Equations (\ref{eq:fvm_formulation_integral_dt}) are evolution equations for the zone-averaged conserved variables $\bar{U}_{n+1}$, which are in turn (non-linear) functions of the zone-averaged primitive variables $\bar{P}_{n+1}$, i.e., $\bar{U}_{n+1} \equiv U(\bar{P}_{n+1})$.

To proceed, we need to evaluate the spatial integrals $\int dv$ and the temporal integrals $\int dt$ in (\ref{eq:fvm_formulation_integral_dt}) using a numerical quadrature to a desired order. We opt for a \emph{truncation error} of $O(\Delta t^2, \Delta X_i^2)$. The required accuracy can be achieved by evaluating the spatial integrals as $\int dX^1 (.) \rightarrow \Delta X^1 (.)_{i}$, $\int dX^2 (.) \rightarrow \Delta X^2 (.)_{j}$, and $\int dX^3 (.) \rightarrow \Delta X^3 (.)_{k}$ where the spatial integer indices $i$, $j$, and $k$ indicate the zone centers in the  $X^1$, $X^2$ and $X^3$ directions respectively. The outcome of this quadrature procedure is that the cell-averaged conserved variables $\bar{U}$, and the cell-averaged source terms $\bar{S}$ can be replaced by point values $U_{i,j, k}$ and $S_{i,j,k}$ at the center of a grid zone and the face-averaged fluxes $\bar{F}^1$ in the $X^1$ direction can be replaced by point values at the centers of the $right$ and $left$ faces, $\bar{F}^1_{right} \approx F^1_{i+1/2, j, k}$ and $\bar{F}^1_{left} \approx F^1_{i-1/2, j, k}$ respectively. The substitution for the face-averaged fluxes $\bar{F}^2$ in the $X^2$ direction, and $\bar{F}^3$ in the $X^3$ direction, by point values follows on similar lines.

\subsection{Time stepping scheme} \label{section:time_stepping}
The temporal integral $\int dt (.)$ for the various terms in (\ref{eq:fvm_formulation_integral_dt}) is approximated to $O(\Delta t^2)$ using a two-stage \emph{semi-implicit} scheme designed to deal with stiff source terms. Depending on the theory being solved for, the source terms can have spatio-temporal derivatives $S \equiv S(P, \partial_t P, \partial_i P)$\footnote{This is an unconventional  definition of source terms, but it allows us to use a notation that is as closely analogous to non-relativistic fluids as possible.}. We separate these as $S = S^I(P) + S^E(P) + A^t(P)\partial_tP + A^i(P)\partial_iP$, where $S^{I,E}(P)$ denote source terms to be treated
implicitly (I) or explicitly (E), and $A^t(P)$, $A^i(P)$ are the coefficients of
the temporal $\partial_t P$ and spatial derivative terms $\partial_i P$
respectively. The spatial derivative terms, when present in the sources, are
evaluated using slope limited derivatives on a symmetric stencil (currently the
generalized minmod slope using a 3 points stencil, although higher-order schemes
inspired by the WENO5 \citep{WENO5a, WENO5b} and PPM \citep{PPM} methods are
also implemented). The scheme proceeds in two stages:
\begin{itemize}
\item First, we take a \emph{half step} to go from  $P_n \rightarrow P_{n+1/2}$, where the index $n+1/2$ indicates the half time step. The temporal integrals for the fluxes $\partial_iF^i$, for the explicit sources $S^E$, and for the spatial derivative terms in the sources are evaluated explicitly using $\int dt(.) \rightarrow (\Delta t/2) (.)_{n}$, whereas the sources $S^I$ are treated implicitly using $\int dt(.) \rightarrow (\Delta t/2)\left((.)_{n+1/2} + (.)_n \right)$ . This leads to the following discrete form
\begin{align} \label{eqn:half_step}
\frac{U(P_{n+1/2}) - U(P_n)}{\Delta t/2}
+ \frac{F^1_{right}(P_n) - F^1_{left}(P_n)}{\Delta X^1} + ... & = \frac{1}{2}\left(S^I(P_{n+1/2}) + S^I(P_n) \right)  \\ \nonumber + & S^E(P_n) + A^t(P_n)\frac{P_{n+1/2} - P_n}{\Delta t/2} + A^i(P_n)\partial_i P_n
\end{align}

\item Next, we take a \emph{full step} from $P_n \rightarrow P_{n+1}$. The temporal integrals for $\partial_i F^i$, $S^E$, and $A^i(P)\partial_iP$ are evaluated using $\int dt(.) \rightarrow \Delta t (.)_{n+1/2} +O(\Delta t^2)$. This is performed using $P_{n+1/2}$ obtained from the half step. The source terms $S^I$ are treated implicitly using $\int dt(.) \rightarrow \Delta t\left((.)_{n+1} + (.)_n \right) + O(\Delta t^2)$
\begin{align} \label{eqn:full_step}
\frac{U(P_{n+1}) - U(P_n)}{\Delta t}
+ \frac{F^1_{right}(P_{n+1/2}) - F^1_{left}(P_{n+1/2})}{\Delta X^1} + ... & = \frac{1}{2}\left(S^I(P_{n+1}) + S^I(P_n) \right)\\ \nonumber  +& S^E(P_{n+1/2}) + A^t(P_{n+1/2})\frac{P_{n+1} - P_n}{\Delta t} \\ \nonumber + & A^i(P_{n+1/2})\partial_i P_{n+1/2}
\end{align}
\end{itemize}
where $(...)$ denote flux discretizations in $X^2$, and $X^3$, which we have not written for brevity.

The separation between explicit and implicit sources $S^{I,E}$ is problem-dependent. Stiff source terms are treated implicitly, while computationally expensive source terms can be treated explicitly if desired. For additional flexibility, nonlinear source terms can also use a mixed implicit-explicit approach. For example, the extended MHD algorithm has source terms of the form
\begin{equation}
S(P) = \frac{P-P_0(\partial_i P)}{\tau_R[P]}
\end{equation}
where $\tau_R$ is a potentially small damping timescale. In this case, it is
advantageous to treat $P/\tau_R[P]$ implicitly and $P_0$ explicitly. But it is
also preferable to use a consistent damping timescale $\tau_R[P]$ for all terms. Accordingly, for the half time step we use
\begin{equation}
\int dt S(P) = \frac{\Delta t}{2} \left(\frac{P_{n+1/2}+P_n}{2\tau_R [P_{n}]} -
  \frac{P_0(\partial_i P_{n})}{\tau_R [P_{n}]}\right),
\end{equation}
and for the full time step,
\begin{equation}
\int dt S(P) = \Delta t \left(\frac{P_{n+1}+P_n}{2\tau_R [P_{n+1/2}]} -
  \frac{P_0(\partial_i P_{n+1/2})}{\tau_R [P_{n+1/2}]}\right).
\end{equation}
This is easily implemented as long as the implicit source terms $S^I$ have access to $P_n$ during the half step and $P_{n+1/2}$ during the full step.
In practice, for any system of equations, the user is responsible for providing functions $S^I(P,P^E)$, $S^E(P^E)$,..., with $P^E=P_n$ for the half-step and $P^E=P_{n+1/2}$ for the full step. The code then assembles the evolution equations from the discretization described in this section.

Evidently, the above system of equations obtained using a semi-implicit temporal
discretization requires us to solve a set of non-linearly coupled equations for
$P_{n+1/2}$ and $P_{n+1}$ in the half step (\ref{eqn:half_step}), and the full
step (\ref{eqn:full_step}) respectively. Further, the presence of time
derivatives $A^t(P)\partial_tP$ in the source terms implies that we cannot
separately time step the conserved variables $U_n \rightarrow U_{n+1} \equiv
U(P_{n+1})$, and invert them later to obtain $P_{n+1}$, as is usually the case.
The time stepping and the inversion must be done simultaneously. We describe the
algorithm to do this in (\S \ref{section:root_finder}). However, we note that
equations without implicitly coupled source terms are treated explicitly, and do
not require the nonlinear solver.

\subsection{Flux Computation} \label{section:spatial_discretization}

The computation of the face-centered fluxes $F^1_{i-1/2} \equiv F^1_i(P^-_{i-1/2}, P^+_{i-1/2})$, and $F^1_{i+1/2} \equiv F^1_{i+1/2}(P^-_{i+1/2}, P^+_{i+1/2})$ requires two stages: (1) \emph{reconstruction} of the primitive variables from the cell centers $P_{...,i-1,i,i+1,...}$ to the left $P^-_{i-1/2,i+1/2}$, and right $P^+_{i-1/2,i+1/2}$ side of the face centers at $i-1/2, i+1/2$, and (2) a \emph{Riemann solver} to evaluates the fluxes $F^1_{i-1/2,i+1/2}$ given the left $P^-_{i-1/2,i+1/2}$ and the right states $P^+_{i-1/2,i+1/2}$.

\subsubsection{Reconstruction}

The face-centered primitive variables are obtained using a reconstruction
operator $R$. The operator takes as input the values of adjacent zone-centered
primitive variables to construct a polynomial interpolant to a desired order
inside the zone, which is then evaluated at the face-centers. We now describe
the reconstruction procedure in one dimension, along $X^1$. For brevity, we
suppress the $X^2$ and $X^3$ zone indices. Multi-dimensional reconstruction
proceeds by performing the one-dimensional reconstruction separately in each
direction.

For a zone with center $i$, the reconstruction operator $R$ is used in two ways depending on the input order. In the case of a 3-point reconstruction stencil, we use $R^+_{i} = R(P_{i-1}, P_{i}, P_{i+1})$ to give $P^-_{i+1/2}$, the primitives variables on the left side of the $right$ face of the zone and $R^-_{i} = R(P_{i+1}, P_{i}, P_{i-1})$ to give $P^+_{i-1/2}$, the primitive variables at the right side of the $left$ face of the zone. This procedure is repeated for the zone with center $i-1$ with $R^+_{i-1}$ to obtain $P^-_{i-1/2}$, and for the zone with center $i+1$ with $R^-_{i+1}$ to obtain $P^+_{i+1/2}$. We now have the states $P^{-,+}_{i-1/2}$ needed by the Riemann solver to compute the fluxes $F_{i-1/2}\equiv F_{i-1/2}(P^-_{i-1/2}, P^+_{i-1/2})$, and $P^\pm_{i+1/2}$ needed to compute $F_{i+1/2} \equiv F_{i+1/2}(P^-_{i+1/2}, P^+_{i+1/2})$.

\subsubsection{Riemann Solver} \label{section:riemann_solver}

For generic systems of equations, we have to rely on relatively simple Riemann solvers -- at least if we want to avoid numerical computation of the characteristic speeds
and eigenvectors of the evolution system. Here, we rely on either the Local Lax
Friedrich (LLF) flux, or the HLLE flux~\citep{HLLE}.
The LLF and HLLE solvers rely on the knowledge of the fluxes $F^\pm_i$ and the conservative variables $U^\pm_i$ on the right/left side of face $i$. Both are computed directly from the reconstructed primitive variables $P^\pm_i$. For the LLF flux, we also use an estimate of the maximum characteristic speed on face $i$, $c_{\rm max,i} \geq \max{(|c^\pm_{j,i}|)}$,
where $c_{j,i}$ is the $j^{th}$ speed on face $i$. The LLF flux is then
\begin{equation}
F^{\rm LLF}_i = \frac{F^+_i + F^-_i}{2} - \frac{c_{\rm max,i}}{2} (U^+_i - U^-_i).
\end{equation}
Similarly, the HLLE flux relies on estimates of the maximum left-going and right-going characteristic speeds on face $i$, $c^R_{\rm max,i} \geq \max{(c^\pm_{j,i},0)}$
and $c^L_{\rm max,i} \geq \max{(-c^\pm_{j,i},0)}$. The HLLE flux is then
\begin{equation}
F^{\rm HLLE}_i = \frac{c^R_{\rm max,i} F^-_i + c^L_{\rm max,i} F^+_i - c^R_{\rm max,i}c^L_{\rm max,i} (U^+_i - U^-_i)}{c^L_{\rm max,i}+c^R_{\rm max,i}}.
\end{equation}
The HLLE flux is generally less dissipative than the LLF flux in regimes where $v^i \gtrsim c_{\rm max}$. The two are identical when the maximum left-going and right-going speeds are equal, but the HLLE flux
smoothly switches to upwind reconstruction when all characteristic speeds have the same sign (e.g., for ideal hydrodynamics, when the speed of the flow across face
$i$ is supersonic). In practice, as the computation of the characteristic speeds for the ideal MHD and EMHD systems can be costly, we replace $c^R_{\rm max}$ and $c^L_{\rm max}$ with simpler analytic upper bounds appropriate for the evolved system of equations.

\section{General Root Finder} \label{section:root_finder}
The spatio-temporal discretization of (\ref{eq:conservative_eqns}) leads to nonlinear equations (\ref{eqn:half_step}) for $P_{n+1/2}$ and (\ref{eqn:full_step}) for $P_{n+1}$. We solve these using an iterative Newton algorithm with a numerical Jacobian assembly, and a backtracking linesearch. The only input to the root finder is a residual function. Thus, we begin by recasting the equations to be solved for, as residuals $R(P)$, where $R$ is the vector of equations, and $P$ are the unknown primitive variables. For example, the residuals for the half step evolution (\ref{eqn:half_step}) are
\begin{align} \label{eqn:half_step_residual}
 R(P_{n+1/2}) & =
 \frac{U(P_{n+1/2}) - U(P_n)}{\Delta t/2} + \frac{F^1_{right}(P_n) - F^1_{left}(P_n)}{\Delta X^1} + ...\\ \nonumber
& -S^E(P_{n}) - \frac{1}{2}\left(S^I(P_{n+1/2}) + S^I(P_n) \right)  - A^t(P_n)\frac{P_{n+1/2} - P_n}{\Delta t/2} - A^i(P_n)\partial_i P_n
\end{align}
Given the residuals as a function of the unknowns $R \equiv R(P)$, the algorithm proceeds by starting with a guess for the unknowns $P_{n+1/2}$ and iterating using
\begin{align}\label{eqn:newton_iters}
P^{k+1} = P^{k} + \lambda^k \delta P^k
\end{align}
for $k=0,1,...,k_{max}$ till $||R(P)|| < tol$, where $||.||$ is a suitable norm,
$tol$ is a desired tolerance, $\delta P^k$ is a linear correction which we
describe in (\S \ref{section:jacobian_assembly}), and $\lambda^k \in (0, 1]$ is
a \emph{linesearch} parameter, which is determined by a quadratic backtracking
linesearch strategy that we describe in (\S \ref{section:linesearch}). In writing the above, we have suppressed the half step index $n+1/2$.

\subsection{Residual-based Jacobian Computation} \label{section:jacobian_assembly}
The correction $\delta P^k$ at each nonlinear iteration $k$ is obtained by solving the following linear system of equations
\begin{align} \label{eqn:numerical_jacobian}
\mathbf{J}(P^k)\delta P^k = -R(P^k)
\end{align}
where the matrix $\mathbf{J}(P^k)$ is the $N_{var} \times N_{var}$ Jacobian of
the system evaluated at $P^k$, and $N_{var}$ is the number of primitive
variables being solved for. The Jacobian itself is assembled numerically from
the residual function $R(P)$ to $O(\epsilon)$, where $\epsilon$ is a small
differencing parameter.  Column $i$, and row $j$ of the numerical Jacobian are
computed using
\begin{align}
\mathbf{J}_{i,j}(P^k) \approx \frac{R_i(P^k_\epsilon) - R_i(P^k)}{P^k_{j,\epsilon} - P^k_j},
\end{align}
and the perturbed unknowns $P^k_\epsilon$ are given by
\begin{align}
  P^k_\epsilon = (1 + \epsilon)P^k(1 - \mathrm{small}(P^k)) + \epsilon *
  \mathrm{small}(P^k)
\end{align}
where
\[
  \mathrm{small}(P^k) =
\begin{cases}
1, & |P^k| < 5 \epsilon \\
0, & \text{otherwise}
\end{cases}
\]
The use of the function $\mathrm{small}(P^k)$ in $P^k_\epsilon$ prevents a
division by zero in (\ref{eqn:numerical_jacobian}). In the absence of
$\mathrm{small}(P^k)$, this occurs when any component of $P^k = 0$, leading to $P^k_\epsilon = 0$.

\subsection{Line Search} \label{section:linesearch}
The traditional Newton algorithm is given by (\ref{eqn:newton_iters}), with $\lambda^k=1$. This is however not robust, and can diverge from the solution. When that happens we \emph{backtrack} by choosing $\lambda \in (0, 1]$ according to the following strategy:

\begin{itemize}
\item Initialize $\lambda^k = 1$.
\item If $||R(P^{k} + \lambda^k \delta P^k)|| < ||R(P_k)|| (1-\epsilon_{BT} \lambda^k)$, accept the new guess $P^{k+1} = P^{k} + \lambda^k \delta P^k$ for the primitive variables and exit the linesearch. Otherwise, continue to the computation of a new $\lambda^k$. In {\tt grim}, we set the small parameter $\epsilon_{BT}=10^{-4}$. If this condition is satisfied, we know that the current guess $P^{k} + \lambda^k \delta P^k$ provides at least some improvement over the previous guess $P^k$.
\item Find the new linesearch parameter $\lambda_{\rm new}$ by minimizing the function
\begin{equation} \label{eqn:linesearch_residual}
f(\lambda^k)=||R(P^{k} + \lambda^k \delta P^k)||^2,
\end{equation}
modeling $f$ as a quadratic function of $\lambda$
and using the fact that $df/d\lambda[\lambda=0] = - 2f(0)$ (as $\delta P^k$ is the solution of the linear problem at $P^{k}$).
We then have
\begin{equation}
\lambda^k_{\rm new} = \frac{f(0)}{f(\lambda^k_{\rm old})+(2\lambda^k_{\rm old}-1)f(0)} \lambda_{\rm old}^2.
\end{equation}
We then set $\lambda^k = \lambda^k_{\rm new}$, and go back to the previous step.
\end{itemize}
We note that this procedure is performed separately at each point. 

\section{Extended GRMHD} \label{section:review_EMHD}

The EMHD model \citep{chandra2015} is a one-fluid model of a plasma consisting of electrons and ions. It considers the following number current vector $N^\mu$ for the ions (set to be the same for electrons) and total (electrons+ions) stress-tensor $T^{\mu \nu}$
\begin{align}
N^\mu & = n u^\mu \\
T^{\mu\nu} & = (\rho + u + \frac{1}{2} b^2)u^\mu u^\nu + (P_g + \frac{1}{2} b^2) h^{\mu\nu} - b^\mu b^\nu + q^\mu u^\nu + q^\nu u^\mu + \Pi^{\mu\nu} \label{eq:fullSEtens}
\end{align}
where $n$ is the number density of ions, which is equal to the number density of
electrons, $\rho = (m_i + m_e) n \approx m_i n$ is the total rest mass energy
density, $m_e$ and $m_i$ are the electron and proton rest masses, $u$ is the
total internal energy, $P$ is the total pressure approximated by a Gamma-law
equation of state $P_g = (\gamma-1)u$, $u^\mu$ is a four-velocity whose choice here corresponds to an observer comoving with the number current, also known as the Eckart frame, and $b^\mu$ is a magnetic field four-vector whose components are given by $b^t =  B^i u^\mu g_{i \mu}, b^i =  (B^i + b^t u^i)/u^t$, where the magnetic field 3-vector $B^i = F^{*it}$, and $F^*$ is the dual of the electromagnetic field tensor. The tensor $h^{\mu\nu}$ is the projection operator onto the spatial slice orthogonal to $u^\mu$,
$h^{\mu\nu} = g^{\mu\nu}+u^\mu u^\nu$. The four-vector $q^\mu$ is a heat flux and the tensor $\Pi^{\mu \nu}$ models viscous transport of momentum. The model ignores bulk viscosity and resistivity. The equations governing $\rho$, $u$, and $u^\mu$ are given by the usual conservation equations,
\begin{align}
\nabla_\mu N^\mu & = 0 \label{eq:divN} \\
\nabla_\mu T^\mu_\nu & = 0
\end{align}
Expanding the covariant derivative $\nabla_\mu$ in a coordinate basis,
\begin{align}
\partial_t\left(\sqrt{-g}\rho u^t\right) + \partial_i\left(\sqrt{-g}\rho u^i\right)& = 0 \label{eq:divN_explicit} \\
\partial_t\left(\sqrt{-g} T^t_\nu\right) + \partial_i\left(\sqrt{-g}T^i_\nu \right) & =
                                               \sqrt{-g}T^\kappa_\lambda \Gamma^\lambda_{\nu\kappa} \label{eq:divT_explicit}
\end{align}
where (\ref{eq:divN_explicit}) has been obtained from (\ref{eq:divN}) by scaling with $m_i$.
The equations governing the components of the magnetic field 3-vector $B^i$ are given by the induction equation in the ideal MHD limit
\begin{align}
\partial_t\left(\sqrt{-g}B^i\right) + \partial_j\left(\sqrt{-g}\left(b^j u^i - b^i u^j\right) \right)& = 0. \label{eq:induction_eqn}
\end{align}
The heat flux $q^\mu$ and the shear stress $\Pi^{\mu \nu}$ that appear in the total stress tensor (\ref{eq:fullSEtens}) are
\begin{align}
q^\mu & = q\, \hat{b}^\mu \\
\Pi^{\mu\nu} & = -\Delta P \left(\hat{b}^\mu \hat b^{\nu} - \frac{1}{3} h^{\mu\nu}\right).
\end{align}
where the scalar $q$ is the magnitude of the heat flux that flows parallel to the magnetic field lines and the scalar $\Delta P = P_\perp - P_\parallel$ is the pressure anisotropy i.e., the difference in pressures perpendicular $P_\perp$ and parallel $P_\parallel$ to the magnetic field. The above forms of the heat flux $q^\mu$ and the shear stress $\Pi^{\mu \nu}$ have been derived by assuming that the distribution functions of all species are gyrotropic, which is accurate in the limit that the Larmor radii are much smaller than the system scale. The evolution of $q$ and $\Delta P$ are given by
\begin{align}
\frac{dq}{d\tau} & = -\frac{q - q_0}{\tau_R} - \frac{q}{2} \,  \frac{d}{d\tau} \log\left(\frac{\tau_R}{\chi P^2}\right) \label{eq:qevol} \\
\frac{d\Delta P}{d\tau} & = - \frac{\Delta P - \Delta P_0}{\tau_R} - \frac{\Delta P}{2}\frac{d}{d\tau}\log \left(\frac{\tau_R}{\rho\nu P}\right),
\label{eq:dPevol}
\end{align}
where $d/d\tau = u^\mu \nabla_\mu$ and
\begin{align}
q_0 & \equiv -\rho\chi\bh^\mu (\nabla_\mu \Theta + \Theta a_\mu) \label{eq:q0def} \\
\Delta P_0 & \equiv 3\rho \nu (\bh^\mu \bh^\nu \nabla_\mu u_\mu - \frac{1}{3}. \nabla_\mu u^\mu )\label{eq:dP0def}
\end{align}
Here, $\Theta = P/\rho \equiv kT/m_ic^2$ is the ion  temperature, $a^\mu \equiv u^\nu \nabla_\nu u^\mu = u^\nu\partial_\nu u^\mu + \Gamma^\mu_{\nu \kappa} u^\nu u^\kappa$ is the four-acceleration and $\chi, \nu$ are the ion thermal and viscous diffusion coefficients respectively. The equations for the heat flux $q$ and the pressure anisotropy $\Delta P$ are obtained by enforcing the second law of thermodynamics. The result then is that $q$ and $\Delta P$ relax to $q_0$ and $\Delta P_0$ over the time scale $\tau_R$, with the additional term in (\ref{eq:qevol}) and (\ref{eq:dPevol}) being of a higher order (if $q \sim \epsilon \ll 1$, then, $qd(\log\left(\tau_R/(\chi P^2) \right))/d\tau  \sim \epsilon^2$ and similarly for $\Delta P$). The terms $q_0$ (\ref{eq:q0def}) and $\Delta P_0$ (\ref{eq:dP0def}) which $q$ and $\Delta P$ relax to respectively, are covariant generalizations of the \cite{Braginskii1965} closure, which the model reduces to in the limit where the relaxation time scale $\tau_R \rightarrow 0$. The above equations (\ref{eq:qevol}) and (\ref{eq:dPevol}) can be rescaled and written
\begin{align}
\nabla_\mu (\tilde{q} u^\mu) &= - \frac{\tilde{q}-\tilde{q}_0}{\tau_R} + \frac{\tilde{q}}{2} \nabla_\mu u^\mu, \label{eqn:q_evol_eqn_rescaled}\\
\nabla_\mu (\Delta \tilde{P} u^\mu) &= -\frac{\Delta \tilde{P} - \Delta \tilde{P}_0}{\tau_R} +
\frac{\Delta\tilde{P}}{2} \nabla_\mu u^\mu, \label{eqn:dP_evol_eqn_rescaled}
\end{align}
with
\begin{align}
\tilde{q} &= q \left(\frac{\tau_R}{\chi \rho \Theta^2}\right)^{1/2}\label{eq:q}\\
\Delta \tilde{P} &= \Delta P \left(\frac{\tau_R}{\nu \rho \Theta}\right)^{1/2}\label{eq:dP}.
\end{align}
These rescaled equations are crucial to our numerical implementation. Equations
(\ref{eq:qevol}) and (\ref{eq:dPevol}) have higher order terms
$q/2\,d(\log(\tau_R/(\chi P^2)))/d\tau$ and $\Delta P/2\, d(\log(\tau_R/(\rho
\nu P)))/d\tau$ which we find are numerically difficult to handle in low density
regions. If these terms are ignored, the positivity of entropy production is no
longer guaranteed. However, the rescaled equations
(\ref{eqn:q_evol_eqn_rescaled}), and (\ref{eqn:dP_evol_eqn_rescaled}) do include
these terms, and using these equations guarantees adherence to the second law of
thermodynamics (up to truncation error in the numerical solution), as well as leads to well behaved numerical solutions.

To conclude, the model evolves, (1) the ion rest mass density $\rho$, (2) the total internal energy density $u$, (3) the spatial components $u^i$ of the four-velocity $u^\mu$, (4) the components of the magnetic field three-vector $B^i$, (5) the ion heat flux along magnetic field lines $q$ and (6) the ion pressure anisotropy $\Delta P$, for a total of ten variables. The governing equations are the continuity equation (\ref{eq:divN_explicit}) for $\rho$, the energy and momentum conservation equations (\ref{eq:divT_explicit}) for $u$ and $u^i$ respectively, the induction equation (\ref{eq:induction_eqn}) for $B^i$, and the relaxation equations (\ref{eqn:q_evol_eqn_rescaled}), and (\ref{eqn:dP_evol_eqn_rescaled}) for $q$ and $\Delta P$ respectively. The inputs to the model are the transport coefficients $\chi$, the thermal diffusivity, and $\nu$, the kinematic viscosity. A closure scheme for $\chi$ and $\nu$ as a function of the relaxation time scale $\tau_R$ is described in \cite{chandra2015}.
The scheme accounts for the presence of kinetic plasma instabilities at subgrid scales, which are prevalent in weakly collisional/collisionless plasmas.
In that closure, we set $\chi = \phi c_s^2 \tau_R$ and $\nu=\psi c_s^2 \tau_R$,  where $\phi$, and $\psi$ are non-dimensional numbers $\sim 1$, $c_s$ is the sound speed, and the damping timescale $\tau_R$ models the effective collision timescale for ions due to kinetic plasma instabilities. 

\subsection{Wave Speeds}

The approximate Riemann solvers allowing us to capture shocks in {\tt grim} require at least an upper bound on the characteristic speeds (See \S \ref{section:riemann_solver}). The speeds control the amount of numerical dissipation introduced in the evolution. To minimize
numerical dissipation, the speed estimates should be as close as possible to the true characteristic speeds, but for stability the estimates should be an
upper bound on the true speeds. In ideal hydrodynamics, the characteristic speeds of the system are known analytically, but this is no longer the case for even ideal
MHD. The characteristic speeds of our EMHD model can be found numerically, but this requires finding the largest and smallest zeroes of a $10^{th}$ degree polynomial on both sides of every cell face. To avoid this expensive operation, we instead follow the methods often implemented in the ideal MHD simulations which use the HLLE or LLF Riemann solvers, and consider an upper bound on the maximum wave speed in the fluid frame, $v_{\rm max}$. We can then obtain upper bounds on the maximum right-going and left-going wave speeds by computing the grid-frame velocity of waves propagating at $\pm v_{\rm max}$ in the rest frame of the fluid, in the direction along which the flux is being computed. A more detailed discussion of the wave speeds of the EMHD model is provided in \cite{chandra2015}. Here we will only note that we use
the practical upper bound
\begin{equation}
v_{\rm max}^2 = \tilde c_s^2 + v_A^2 - \tilde c_s^2 v_A^2
\end{equation}
where $v_A$ is the usual Alfven speed and $\tilde c_s$ is a correction to the sound speed including the effects of heat conduction and viscosity:
\begin{align}
v_A^2 &= \frac{b^2}{\rho +\gamma u + b^2}\\
c_s^2 &=\frac{\gamma(\gamma-1)u}{\rho+\gamma u}\\
\tilde c_s^2 &= \frac{1}{2}\left( c_s^2 + v_q^2 + \sqrt{c_s^4 + v_q^4}\right) + v_{\Delta P}^2. \label{eq:modcs}
\end{align}
The corrections $v_q$ and $v_{\Delta P}$ to the sound speed $c_s$ are
\begin{align}
v_q^2 &= (\gamma-1)\frac{\chi}{\tau_R},\\
v_{\Delta P}^2 &= \frac{4\nu}{3\tau_R}.
\end{align}
With the closure scheme in \citep{chandra2015}, the speed $\tilde c_s$ simplifies to
\begin{equation}
\tilde c_s^2 = \frac{c_s^2}{2} \left(1 + (\gamma-1)\phi + \sqrt{(1 + (\gamma-1)^2\phi^2 } +\frac{8\psi}{3}\right).
\end{equation}
From this equation and the inequality $c_s^2 \leq (\gamma-1)$, we can also derive conditions on $\psi$ and $\phi$ which guarantee $v_{\rm max}^2<1$.
This is a sufficient, but not necessary, condition for the system to be causal and hyperbolic. For $\psi=\phi$ and the standard choice of
$\gamma=4/3$ (resp. $5/3$), we find $\psi_{\rm max}\approx 1.3$ (resp. $\psi_{\rm max}\approx 0.29$). As, at the level of the Riemann solver, we do not assume a specific closure scheme in {\tt grim}, we implement Eq.~\ref{eq:modcs} for $\tilde c_s$, and not the simplified version.

\subsection{Constrained Transport}
A crucial ingredient for the evolution of the induction equation (\ref{eq:induction_eqn}) is the preservation of the zero monopole constraint $\nabla\cdot \mathbf{B} = 0$. Naive evolution leads to uncontrolled growth of the constraint, resulting in numerical instabilities. Constrained transport schemes \citep{evans} exactly preserve a specific numerical representation of the constraint, i.e., the violations are at machine tolerance. 

We use a version of constrained transport by \citealt{toth}, the \emph{flux-CT}
scheme, where the magnetic fields are co-located with the fluid variables, at
the cell-centers\footnote{We are currently testing a version that uses face
centered formulation.}. They then are evolved by the same routines in a finite
volume sense, with the ``fluxes'' being the electric fields (up to a sign),
which for the EMHD model (just as in ideal MHD) are $F^j = \sqrt{-g} \left(b^j
u^i - b^i u^j\right)$. At the end of the update the face centered fluxes
$F^j_{\rm face}$ (i.e. the electric fields) obtained from the Riemann solver are
averaged to the edges to get the edge centered fluxes $F^j_{\rm edge}$. The edge
centered fluxes are then averaged to get new face centered fluxes
$\bar{F}^j_{\rm face}$, which are then used to evolve the volume averaged
magnetic fields $\int B^i \Delta v$. The simple averaging procedure we use
$F^i_{\rm face} \rightarrow F^i_{\rm edge}$ is the original \citealt{toth}
formulation, which is also being used in the {\tt harm} code, and lacks
upwinding information (see \citealt{gardiner} for a discussion on the
limitations of this approach). 

\section{Implementation Details} \label{section:implementation}
We now discuss the implementation of the algorithms described in the previous sections. {\tt grim} is written in C++, with a modular library architecture. Different components such as spatial reconstruction, the Riemann solver, boundary conditions, and evaluation of the metric and related quantities are are all separate libraries. Each library has automated unit tests to ensure robustness against inadvertent programmer errors \footnote{At present, 75 units tests.}.

{\tt grim} is designed to run on existing, as well as upcoming architectures. It
has been tested and benchmarked on CPUs as well as on Nvidia and AMD GPUs. In
(\S \ref{section:performance}), we formulate a performance model, and describe the specifications of a machine that {\tt grim} is most sensitive to. Guided by the model, we have optimized {\tt grim} to achieve a significant fraction of machine  peak on both CPU and GPU systems.

\subsection{Dependencies}

{\tt grim} is built on top of the {\tt PETSc} \citep{petsc} library to handle distributed
memory parallelism, and the {\tt ArrayFire} \citep{Yalamanchili2015} library for
shared memory parallelism within a node. The C++ vector abstractions from {\tt
ArrayFire} allow {\tt grim} to run on a variety of computer architectures (CPUs and GPUs) using the \emph{same} code. We discuss this in detail in
(\S \ref{section:parallelism}) and then describe how we integrate {\tt PETSc} and
{\tt ArrayFire} to achieve architecture agnostic distributed memory parallelism
in (\S \ref{section:grid}).

\subsection{Architecture Agnostic Code} \label{section:parallelism}
There are now several supercomputers \footnote{\url{www.top500.org}} that, in addition to CPUs, have accelerators such as GPUs. The programming models for these two architectures are different. We are able to write a single code that runs on both architectures by performing operations within a node using the {\tt array} data structure from the {\tt ArrayFire} library. Operations to be performed on an {\tt array} are written down in a vector notation. For example, to add {\tt array A}, {\tt array B}, and write to {\tt array C}, each of which hold multidimensional data,  we write {\tt C = A + B} to perform the operation over the entire domain. At runtime, {\tt ArrayFire} detects the available compute architectures on the node, and fires kernels customized to that architecture, using either an {\tt OpenCL}, {\tt CUDA}, or {\tt CPU} backend.

The use of vector notation also significantly simplifies the code. The entirety of our implementation of the nonlinear solver \S \ref{section:root_finder}, including the Jacobian assembly \S \ref{section:jacobian_assembly}, the linear inversion (\ref{eqn:numerical_jacobian}), and the quadratic backtracking linesearch \S \ref{section:linesearch} is 250 lines (including comments).

All the mathematical operations that need to be performed in {\tt grim} can be divided into two categories, \emph{local} operations that operate point-wise, and \emph{non-local} operations that require data from adjacent grid zones, such as reconstruction. We describe how both of these are implemented using the vector notation.

\subsubsection{Local Operations}
A majority of calculations in {\tt grim} such as computing the conserved
variables $U(P)$, the fluxes $F^{1,2,3}(P)$, and the various source terms $S^{I,
E}(P)$, $S(\partial_tP)$ involve point-wise operations. These are easily
implemented in vector notation, with certain caveats. As will be described in
(\S \ref{section:performance}), the speed of vector-vector operations is set completely by the available memory bandwidth of the system, and therefore it is crucial to maximize the \emph{effective} bandwidth.

We illustrate what effective bandwidth means with the following computation: we have a contravariant four-vector $u^\mu$, that we want to transform to a covariant four-vector $u_\mu$, using $u_\mu = g_{\mu \nu} u^\nu$, where $g_{\mu \nu}$ is the metric. Converting to computer code in vector notation, we have
\lstset{basicstyle=\ttfamily,breaklines=true}
\begin{lstlisting}
for (int mu=0; mu < 4; mu++)
{
  uCov[mu] = 0;
  for (int nu=0; nu < 4; nu++)
  {
    uCov[mu] += gCov[mu][nu]*uCon[nu];
  }
}
\end{lstlisting}
Each of {\tt uCov[mu], uCov[mu], gCov[mu][nu], uCon[nu]} is an {\tt array} of size $N_{1} \times N_{2} \times N_{3}$, where $N_{1}$, $N_{2}$, and $N_{3}$ are the number of grid zones in the $X^1$, $X^2$, and $X^3$ directions respectively, on each node. The operation {\tt uCov[mu] += gCov[mu][nu]*uCon[nu]}, occurs over all $N_{1} \times N_{2} \times N_{3}$ grid zones. 

Notice that {\tt uCon[nu]}, {\tt nu = 0, 1, 2, 3}, is being read for the computation for each of {\tt uCov[mu]}, {\tt mu = 0, 1, 2, 3}. For grid sizes that exceed the cache, as is the case with production science runs, this involves reads from slow global memory and is therefore a performance bottleneck. In this case, the computation of {\tt uCov[mu]} involves 32 global reads (16 for {\tt gCov[mu][nu]}, and 4$\times$4=16 for {\tt uCon[nu]}). An optimal implementation would involve 20 global reads, with only 4 reads for {\tt uCon[nu]}. Therefore, the effective bandwidth achieved is only $20/32 \approx 0.62$ of the ideal value. Thus, while the abstraction of mathematical operations using vector notation allows for computation to be performed on a wide variety of computer architectures, further innovation is required to ensure optimality of the computation. 

The feature that enables near-optimal performance of point-wise vector
computations is {\tt ArrayFire}'s \emph{lazy} evaluation using its Just-In-Time
(JIT) compiler. To avoid multiple reads of the memory, the operations that need
to be performed on the arrays {\tt uCov[mu]} are queued, instead of being
immediately executed (known as \emph{eager} evaluation, as is usually the case).
Execution occurs using {\tt eval(uCov[0], uCov[1], uCov[2], uCov[3])}. The JIT
analyses the common dependencies between all four arrays, and fires a single
kernel without any redundant reads and writes. In the above example, this leads
to a single read for {\tt uCon[mu]} instead of four separate reads. Our
measurements indicate that this leads to architecture independent optimal
effective bandwidth, which is crucial to the performance of our nonlinear
solver. We discuss this further in (\S \ref{section:performance}).

\subsubsection{Non-local Operations}

Operations such as reconstruction and interpolation can be thought of as
non-local operations because they operate on stencils of non-zero width, as
opposed to point-wise local operations that operate on stencils of zero width.
Non-local operations are performed using discrete convolutions. The abstraction
of finite differences as discrete convolutions has two advantages: (1) it allows
for architecture agnostic code since all we require is an optimized convolution
routine for CPUs and GPUs, and (2) there are indeed optimized convolution
routines for both these architectures because convolutions are crucial to image
processing.

A discrete convolution of input data $g$, with a filter $f$, at a point $n \in [0, N)$ is defined as
\begin{align}
(f * g)[n] \equiv \sum\limits_{m=-M}^M f[m] g [n-m]
\end{align}
where $g$ is an {\tt array} of size $N$, and the filter $f$ has a stencil width $2M+1$ with extent $\{-M, -M+1,...,0,..., M-1, M\}$. Forward differences $dg_+[n] \equiv g[n+1] - g[n]$ are computed using $f = \{1, -1, 0\}$, while backward differences $dg_-[n] \equiv g[n] - g[n-1]$ are computed using $f = \{0, 1, -1\}$. Central differences $dg \equiv g[n+1] - g[n-1]$ are simply $dg = dg_+ + dg_-$.

We use the optimized {\tt convolve()} function provided by {\tt ArrayFire} that takes in an input {\tt array g} of size $N$, and a set of $P$ filters ${\tt f_1, f_2,..f_P}$, and simultaneously operates all filters over the input data to return an {\tt array h} with dimensions $N\times P$. The {\tt array h} then contains the forward $dg_+$, and backward differences $dg_-$ along a specified direction, over the entire domain. The combination of these two with vectorized conditional operators such as {\tt c = min(a, b)} allows us to implement the slope limiters that are required for the reconstruction operation.

\subsection{Parallelization Infrastructure} \label{section:grid}
One of the many mundane tasks involved in writing a finite volume code is the allocation of memory and initialization of several $N_1 \times N_2 \times N_3 \times N_{var}$ arrays, where $N_1$, $N_2$, and $N_3$ are the number of grid zones along $X^1$, $X^2$ and $X^3$ directions respectively, and $N_{var}$ is the number of variables at each grid zone.

In addition to the memory allocation, there are several other functions the code needs for (1) partitioning the data across several nodes in a distributed memory cluster, (2) communication of ghost zones between nodes that share the same boundary, and (3) parallel file input and output that works with data spread over several nodes. To do all of the above, we created the {\tt grid} class which forms the backbone of {\tt grim}.

\subsubsection{Parallelization}
An instance of the {\tt grid} class is created using {\tt grid prim(N1, N2, N3, Ng, dim, Nvar)} \footnote{For the exact form of the definitions, please refer to the source code}, where {\tt N1}, {\tt N2}, and {\tt N3} are the number of grid zones along $X^1$, $X^2$ and $X^3$ directions respectively, {\tt Ng} is the number of ghost zones required, {\tt dim} is the dimension, and {\tt Nvar} is the number of variables at each zone. This builds a structured grid, performs domain decomposition using {\tt PETSc} over a chosen number of distributed nodes, and creates {\tt prim.vars[0], prim.vars[1], ..., prim.vars[Nvar-1]}, each of which is an {\tt array} from the {\tt ArrayFire} library which lives on either CPUs, or GPUs, depending on the node architecture.

Each {\tt array} is a contiguous block of memory of size {\tt (N1Local + Ng)} $\times$ {\tt (N2Local + Ng)} $\times$ {\tt (N3Local + Ng)}, where {\tt N1Local, N2Local} and {\tt N3Local} are the local sizes of the domain on each node. This arrangement of variables in memory is known as Struct of Arrays (SoA), leading to vectorized pointwise operations, and contiguous memory accesses. This results in optimal memory bandwidth usage, which as we discuss in \S \ref{section:performance} determines {\tt grim}'s performance.

Communication of ghost zones is performed by simply calling {\tt prim.communicate()}, which will exchange ghost zone data of all {\tt Nvar} variables in {\tt prim} using {\tt MPI}. The {\tt communicate} function works independent of where the data lies, whether on the host CPU or attached GPU(s). If the data is on GPUs, it is transferred to the host, the ghost zone data is exchanged, and transferred back to the GPUs.

\section{Performance and Scaling} \label{section:scaling}

\begin{figure}[!htbp]
\begin{center}
\includegraphics[width=140mm]{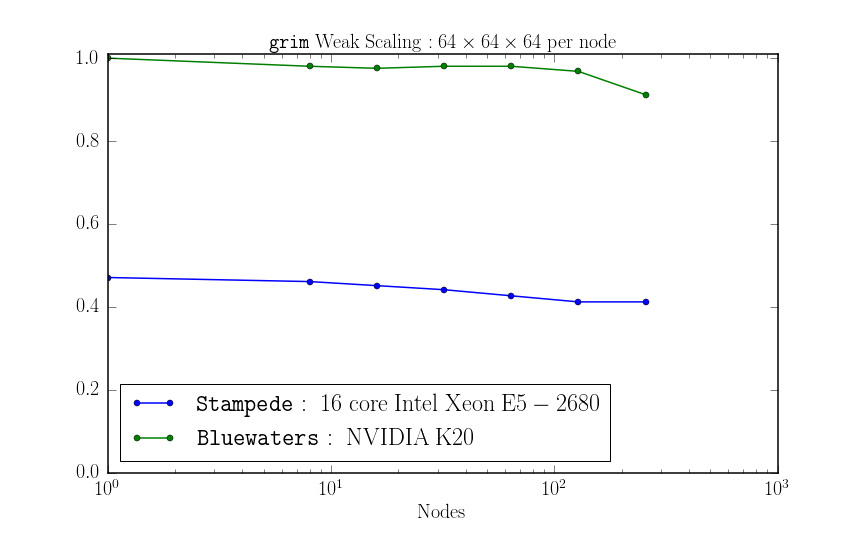}
\end{center}
\caption{Weak scaling on 256 CPU nodes (4096 cores) of Stampede, compared to 256 GPU nodes of Bluewaters. The performance has been normalized to the value of that on a single Bluewaters node. The scaling on both machines is $\sim 93 \%$.}
\label{fig:scaling}
\end{figure}

We have benchmarked {\tt grim} on clusters with varied architectures. On the
Stampede supercomputer, using NVIDIA K20 GPUs, {\tt grim} evolves $138,000$ grid zones/sec/GPU, with $64 \times 64 \times 64$ zones per GPU. On the CPU nodes which have a 16 core (2 sockets x 8 cores each) Intel Xeon E5-2680 CPU, and the same resolution per node, the performance is $48,000$ zones/sec/node. {\grim} scales well on both CPU  and GPU machines. Fig.~\ref{fig:scaling} shows $\sim 93 \%$ weak scaling up to 4096 CPU cores on Stampede, and 256 GPUs on Bluewaters.

The primary difference in speed when using {\tt grim} on GPUs, as compared to CPUs is due to the higher memory bandwidth available on GPUs. The typical accessible bandwidths on GPUs are $\sim 140$ GB/s, while on CPUs it is $\sim 50$ GB/s (using all cores on all sockets). Based on this, we expect a single GPU to be $\sim 2-3$ times faster than a multicore CPU for our implementation.

\section{Performance model} \label{section:performance}

The performance numbers quoted in \S \ref{section:scaling} are
\emph{experimental} results. They do not give information regarding the
efficiency of our implementation of the algorithms described in (\S
\ref{section:finite_volume_method} and \S \ref{section:root_finder}). In order to do so, we require a performance model to benchmark against.

Performance of a code on a given machine is broadly set by two factors, the algorithm, and the implementation. Our root finder (\S \ref{section:root_finder}) while allowing for the exploration of a broad range of fluid-like theories is much slower (factor of $\sim 50$ slower per CPU core) than the schemes \citep{primToCons} used for primitive variable inversion in ideal relativistic fluids. As a result, the dominant cost in {\tt grim} ($\sim 90 \%$) is the nonlinear solver, with the sum of the reconstruction procedure, and the Riemann solver taking only $\sim 5 \%$ of the time. Therefore, we focus our efforts on understanding the costs involved in the nonlinear solver.

The nonlinear solver involves the following three steps (1) Jacobian assembly, (2) solution of a block diagonal linear system, and (3) linesearch. For the linear solver, we use vendor provided LAPACK routines, which we assume are already optimized. Therefore, we only consider the Jacobian assembly and the linesearch, both of which are performed by repeated calls to the residual function $R(P)$. Given a guess for the $N_{\rm var}$ primitive variables $P$, the $N_{\rm var}\times N_{\rm var}$ Jacobian $J(P)$ (eq. \ref{eqn:numerical_jacobian}) is assembled using $N_{\rm var}$ calls to the residual function $R(P)$ (eq. \ref{eqn:half_step_residual}) that returns a vector of size $N_{\rm var}$. Similarly, the linesearch algorithm only depends on the residual function, through it norm $f(\lambda) = ||R(P + \lambda \delta P)||^2$ (eq. \ref{eqn:linesearch_residual}). Thus, it is sufficient to analyze the operations involved in the residual function.

\subsection{Residual assembly}

Consider assembly of the residual $R(P_{n+1/2})$.  It is assembled with calls to functions that compute the conserved variables $U(P)$ and the source terms $S^I(P)$, $S(\partial_tP) \equiv A^t(P_n)(P_{n+1/2} - P_n)/(0.5\Delta t)$, each of which return a vector of size $N_{\rm var}$. The other terms in the residual, $U(P_n)$, $F^{1,2,3}(P_n)$, $S^I(P_n)$, and $A^i(P_n)\partial_iP_n$, only involve $P_n$, the primitive variables at the previous time step, and are precomputed outside the residual assembly. Therefore, the performance of the residual computation, and hence the Jacobian assembly, and the linesearch are set by the performance of the functions to compute $U(P)$, $S^I(P)$, and $S(\partial_tP)$. We now discuss the main factor that determines the runtime of these functions, the memory bandwidth of the system.

\subsubsection{Primary Architectural Bottleneck}

Consider the computation of the fluid conserved variables $U(P) \equiv T^0_\nu$, with an ideal MHD stress tensor, for brevity. The computation requires the density $\rho \equiv \mathtt{rho}$, internal energy $u \equiv \mathtt{u}$, pressure $P \equiv \mathtt{P}$, the four-velocities $u^\mu \equiv \mathtt{uCon}$, $u_\mu \equiv \mathtt{uCov}$, the magnetic field four-vectors $b^\mu \equiv \mathtt{bCon}$, $b_\mu \equiv \mathtt{bCov}$, and the magnetic pressure $b^2 \equiv \mathtt{bSqr}$,
\begin{lstlisting}
for (int nu=0; nu < 4; nu++)
{
  T[0][nu] =   (rho + u + P + bSqr)*uCon[0]*uCov[nu]
              + (P + bSqr/2)*delta(0, nu)
              - bCon[0]*bCov[nu];
}
\end{lstlisting}
where $\mathtt{delta(0, nu)} \equiv \delta^0_\nu$ is the Kronecker delta.

The above code has a total of 11 floating point operations, 14 reads {\tt rho, u, P, bSqr, uCon[0], uCov[nu], bCon[0], bCov[nu]}, and four writes {\tt T[0][nu]}. The total time taken to execute the above code is the time taken to load the data, perform the floating point operations, and finally write the data. Therefore, the total time taken is
\begin{align}
t_{total} &= (N_{reads}t_{read} + N_{flops}t_{flops} + N_{writes}t_{write})N
\end{align}
where $N_{reads}$, $N_{writes}$, and $N_{flops}$ are the total number of reads,
writes, and flops performed per grid zone, $N$ is the total number of grid zones, and
$t_{read}$, $t_{write}$, and $t_{flops}$ is the time taken by the machine to
perform a single read, write, and a floating point operation respectively. The
parameters $t_{read}$, $t_{write}$, and $t_{flops}$ are architecture  and
machine specific. The specifications are usually given in terms of floating
point operations per second \emph{flops}, and memory bandwidth \emph{Bytes/sec}.
Typical peak numbers for a current CPU are 500 \emph{Gflops}, and 100
\emph{GB/sec}. For $N \sim 10^9$ (and hence ignoring latency effects), these
correspond to $N t_{flops} \sim 0.02$ seconds, $N t_{read} \sim 1.12$ seconds,
and $N t_{write} \sim .32$ seconds. Evidently, the ratio $(t_{reads} + t_{writes}) / t_{flops} \gg 1$. Therefore, the runtime of the above code is almost completely set by how fast the data can be transferred between the memory system and the compute units. The actual computation time is negligible, as long as $N_{flops}/(N_{reads} + N_{writes}) \sim 1$, which is indeed the case for all functions involved in the Jacobian assembly and the linesearch.

\subsubsection{Effective Bandwidth Usage}

Since the performance is set by the speed of memory access, we can calculate the time it should take to compute the functions $U(P)$, $S^I(P)$, and $S(\partial_tP)$ by simply examining the inputs $N_{\rm reads}$, and the outputs $N_{\rm writes}$ to each function. The calculation is independent of the exact operations $P \rightarrow \{U(P), S^I(P), S(\partial_tP)\}$, and is given by
\begin{align} \label{eqn:bandwidth}
t\; ({\rm secs}) & = \frac{(N_{\rm reads} + N_{\rm writes}) \times 8}{10^9} \times \frac{1}{{\rm Bandwidth\;(GB/sec)}}
\end{align}
where $N_{\rm reads}$, and $N_{\rm writes}$ are the number of reads, and writes of double precision variables, each of which are 8 bytes. By measuring the runtime $t$ of each function, the \emph{effective} bandwidth being used is calculated using (\ref{eqn:bandwidth}). 

The measured bandwidth used in each function is now normalized with that obtained from the {\tt STREAM} benchmark, given by the operation {\tt c = a + b}, where {\tt a}, {\tt b}, and {\tt c} are arrays of sizes equal to the local grid sizes after domain decomposition. The {\tt STREAM} benchmark has $N_{\rm reads} = 2$ ({\tt a}, {\tt b}), $N_{\rm writes} = 1$ ({\tt c}), and is a metric of the sustained bandwidth that can be obtained on a given machine. The typical value of this benchmark on GPUs is $\sim 140$ GB/sec, whereas on CPUs it is $\sim 50$ GB/sec for array sizes that exceed the cache, and when using all cores on all sockets \footnote{Comparing a single CPU core to an entire GPU is not representative of how CPUs are used in production runs. Using a single core of a CPU leads to bandwidths that are much lower than the peak. In order to saturate the bandwidth, it is \emph{necessary} to use $\gtrsim 50\%$ of all available cores.}. These numbers inform us about the potential speedup of bandwidth limited operations on GPUs, compared to CPUs.

By comparing the measured bandwidth of each function to the bandwidth obtained
from the {\tt STREAM} benchmark, we get the efficiency of our implementation,
which we find is $\sim 70-80 \%$ on both GPUs and CPUs. A significantly lower
($\lesssim 20 \%$) value indicates that there are either superfluous memory
accesses that are not accounted for, or non-contiguous
memory accesses that are not vectorized. Both of these reduce the effective
memory bandwidth. The high bandwidth obtained by our implementation indicates
that we have accounted for leading order performance bottlenecks in the residual
evaluation, leading to a near-optimal Jacobian assembly  and linesearch.

\section{Test Suite} \label{section:tests}

{\tt grim} has been tested extensively in the linear, nonlinear, special
and general relativistic regimes. The tests below are grouped according to the
physical model being solved, with subsections describing individual tests.

\subsection{Extended MHD}

\subsubsection{Linear modes}
An important check of any numerical implementation of the EMHD model is whether it
can reproduce the corresponding linear theory with an error that falls off at the
expected order of spatio-temporal discretization. In order to perform this
test, one needs the linear theory of the EMHD model.

The governing equations of EMHD are considerably more complicated than the governing equations of ideal MHD.
In particular the inclusion of both anisotropic pressure
and conduction, which are sourced by spatio-temporal derivatives projected along
the magnetic field lines, make it challenging to derive the linear theory; the derivation is prone to errors if done manually.  To address this issue, we have
written a general linear analysis package \footnote{{\tt balbusaur}:
\url{http://bit.ly/2bEGW4l}} built on top of the \cite{sage}  computer algebra system, which takes
as input the governing equations of any model, and generates the characteristic
matrix of the corresponding linear theory. The eigenvectors of this matrix are
then used as initial conditions in {\tt grim}, and their numerical
evolutions checked against the corresponding analytic solutions.

\begin{center}
    \begin{tabular}{ l || l | clc } \label{table:eigenvector}
Variable ($P$) & Background State ($P_0$) & Perturbed Value ($\delta_P$) \\ \hline \hline
$\rho$   &     1.           & $-0.518522524082246 - 0.1792647678001878i$  \\
$u$      &     2.           & $0.5516170736393813$  \\
$u^1$    &     0.           & $0.008463122479547856 + 0.011862022608466367i$  \\
$u^2$    &     0.           & $-0.16175466371870734 - 0.034828080823603294i$          \\
$u^3$    &     0.           & 0.          \\
$B^1$    &     0.1          & $-0.05973794979640743 - 0.03351707506150924i$     \\
$B^2$    &     0.3          & $0.02986897489820372 + 0.016758537530754618i$          \\
$B^3$    &     0.           & 0.          \\
$q$      &     0.           & $0.5233486841539436 + 0.04767672501939603i$          \\
$\Delta P$&     0.          & $0.2909106062057657 + 0.02159452055336572i$         \\
    \end{tabular} \\
    \vspace{1mm}
    \text{Table 1: Eigenvector with eigenvalue for EMHD linear modes test.}
\end{center}

Our linear test uses a propagating mode with wave vector $k_1 = 2\pi, k_2 = 4 \pi$ misaligned with the background magnetic field $\mathbf{B}_0 = (0.1, 0.3, 0)$.  Both $\mathbf{k}$ and $\mathbf{B}$ are misaligned with the numerical grid.  We use the eigenvector tabulated in table (\ref{table:eigenvector}). Each of the variables are initialized as $P = P_0 + A\delta_P \exp(i(k_1 x^1 + k_2 x^2))$, where $P$ is the variable, $P_0$ is the background state, $\delta_P$ is the perturbed values, and $A$ is the amplitude of the perturbation, which we set to $10^{-8}$. The exact solution is given by $P = P_0 + A\delta_P\exp(i(k_1 x^1 + k_2 x^2) + \omega t)$, where $\omega = -0.5533585207638141 - 3.6262571286888425i$.  The mode is both propagating and  decaying, indicating the presence of dissipation. 

\begin{figure}[!htbp]
\begin{center}
\includegraphics[width=90mm]{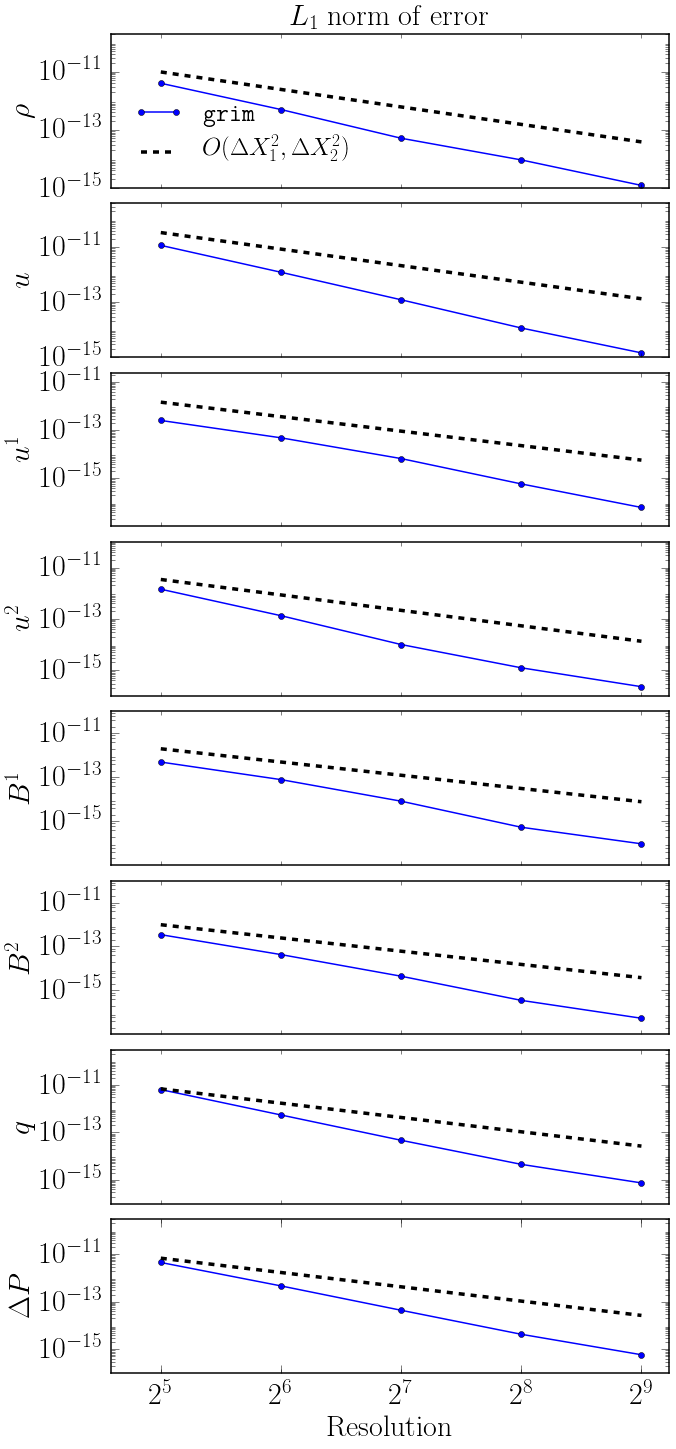}
\end{center}
\caption{Convergence in the linear modes test.}
\label{fig:linear_modes}
\end{figure}

The mode is evolved in a box with dimensions $[0, 1] \times [0, 1]$, periodic
boundary conditions, and  resolutions $(N_1, N_2) = (32, 32), (64, 64), ...,
(512, 512)$. The diffusion coefficients are $\chi = c_s^2 \tau_R$, and $\nu =
c_s^2 \tau_R$, with $\tau_R = 1$, $c_s^2 = \gamma P/(\rho + \gamma u)$, $P_g = (\gamma - 1)u$, and $\gamma = 4/3$. We compare the numerical and analytic solutions at $t = 0.5$. Fig.~\ref{fig:linear_modes} shows that the $L_1$ norm of the error falls off at the expected order.

\subsubsection{EMHD Shock Solutions}

\begin{figure}[!htbp]
\begin{center}
\includegraphics[width=120mm]{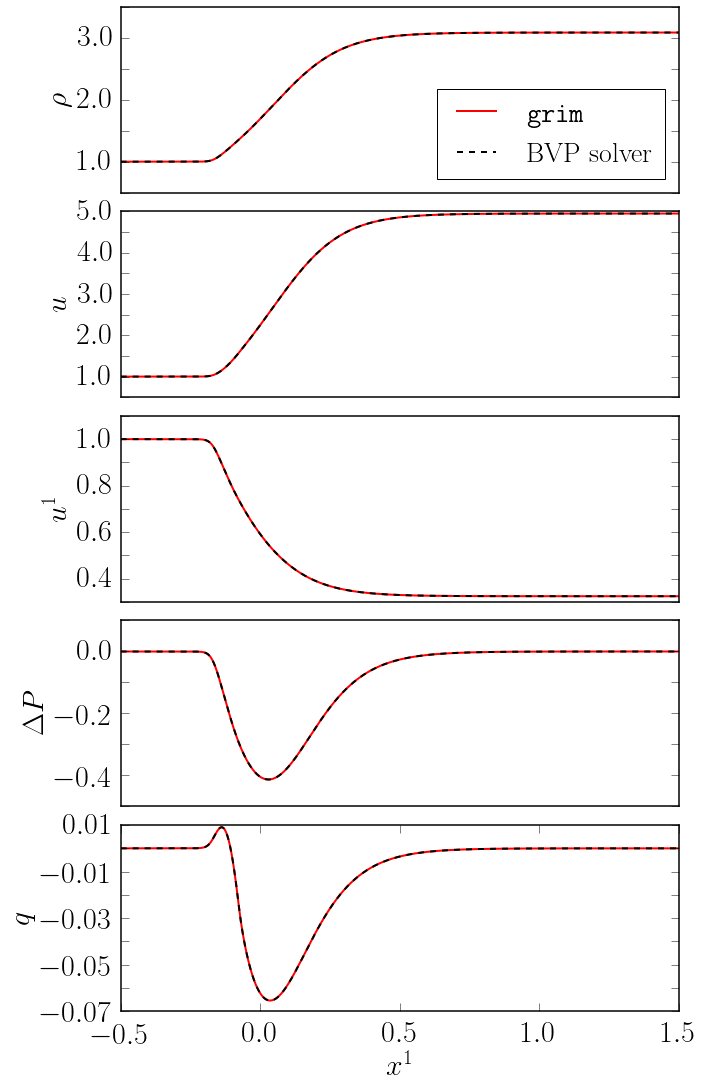}
\end{center}
\caption{Shock solution obtained by {\tt grim}, which solves the EMHD theory as an initial value problem (IVP), plotted on top of the shock solution of the EMHD theory, solved as a boundary value problem (BVP).}
\label{fig:bvp_vs_grim}
\end{figure}

In EMHD, viscosity can smooth a shock and connect the left and right states with a well-defined solution. The hyperbolic nature of the dissipation leads to new features in the shock structure which have been qualitatively described in \cite{chandra2015}. Here, we solve the magnetic field aligned shock structure in the EMHD model as a boundary value problem (BVP) with the left and right states fixed to the values given by the Rankine-Hugoniot jump conditions. We then use this as a reference solution to check the EMHD shock solutions obtained from {\tt grim}, which solves the EMHD equations as an initial value problem (IVP) (fig.~\ref{fig:bvp_vs_grim}). 

The boundary value solutions are obtained using a global Newton root finder. We are looking for a steady state time independent nonlinear solution of the EMHD equations, and hence set the time derivatives $\partial_t \rightarrow 0$. Since we are interested in the \emph{continuous} shock sub-structure, we approximate all spatial derivatives $\partial_x$ by central differences with a truncation error $O(\Delta x^8)$. Thus we have a set of coupled discrete nonlinear equations $R(P_i) = 0$, where $P_i$ are the primitive variables at $i=0, 1, ..., N_x$, and $N_x$ is the chosen spatial resolution of the numerical grid. The system is iterated upon starting from a smooth initial guess using the Newton's method combined with a numerical Jacobian assembled to machine precision. The iterations are continued until we achieve machine precision error $O(10^{-14})$.

\begin{figure}[!htbp]
\begin{center}
\includegraphics[width=120mm]{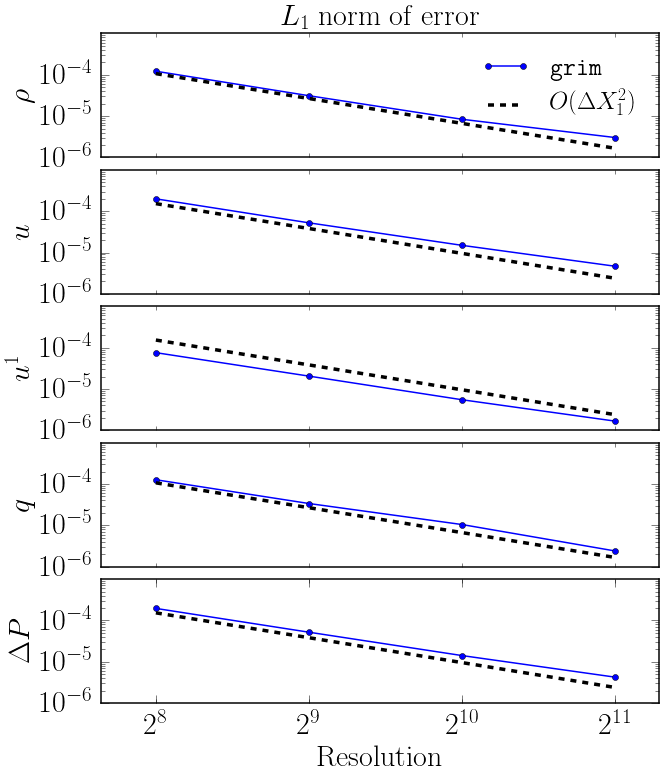}
\end{center}
\caption{Convergence of a resolved EMHD shock, between an initial value problem solved with {\tt grim}, and a boundary value problem solved by setting $\partial_t \rightarrow 0$ in the EMHD theory.}
\label{fig:EMHD_shock_convergence}
\end{figure}

The solution obtained from the initial value problem starting from a discontinuous initial condition (shown in table (\ref{table:emhd_shock})), and the solution obtained from the boundary value problem are connected by a translation. For a quantitative check of the error, we use the BVP solution as an initial condition into {\tt grim}, and check for convergence after a fixed time. Fig.~\ref{fig:EMHD_shock_convergence} shows convergence between the two solutions as a function of resolution.

\begin{center} 
    \begin{tabular}{ l || l | l } \label{table:emhd_shock}
Variable & Left State & Right State \\ \hline \hline
$\rho$     &     1.     & 3.08312999  \\
$u$        &     1.     & 4.94577705  \\
$u^1$      &     1.     & 0.32434571  \\
$u^2$      &     0.     & 0.          \\
$u^3$      &     0.     & 0.          \\
$B^1$      &     $10^{-5}$& $10^{-5}$     \\
$B^2$      &     0.     & 0.          \\
$B^3$      &     0.     & 0.          \\
    \end{tabular} \\
    \vspace{1mm}
    \text{Table 1: Steady state shock solution in Ideal MHD}
\end{center}

\begin{figure}[!htbp]
\begin{center}
\includegraphics[width=140mm]{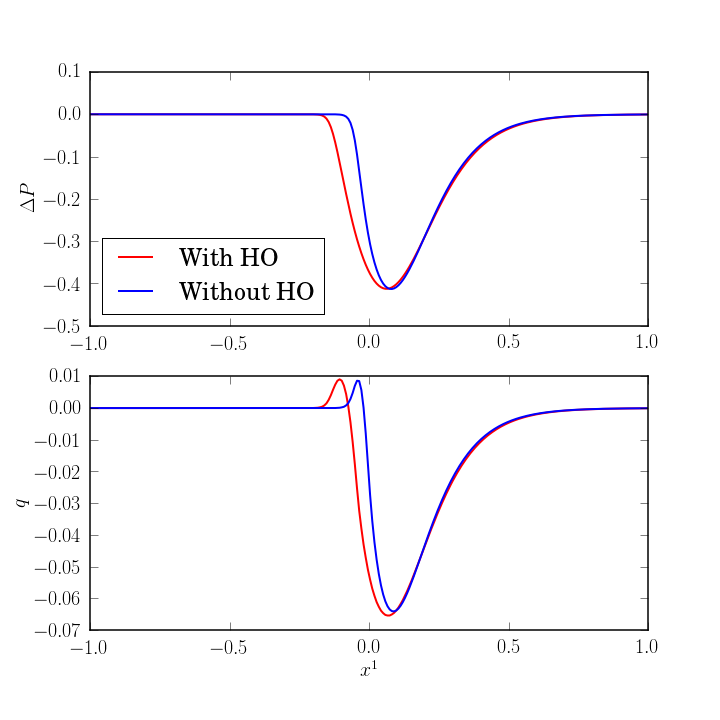}
\end{center}
  \caption{Dependence of the shock substructure on the presence of higher order
  (HO) terms $\sim q u^\mu \nabla_\mu(\tau_R/(\chi P^2)), \Delta P u^\mu
  \nabla_\mu(\tau_R/(\rho \nu P))$ in the EMHD theory.}
\label{fig:HO_vs_NO_HO}
\end{figure}

The EMHD theory has three free parameters which we set to the following values:
the relaxation time scale $\tau_R = 0.1$, the kinematic viscosity $\nu = \psi
c_s^2 \tau_R$, and the thermal diffusivity $\chi = \phi c_s^2 \tau_R$, with the
non-dimensional parameters $\psi = 3$ and $\phi = 5$. To get a continuous shock
solution we require that the characteristic speed of viscosity in the EMHD
theory $v_{\rm char} \sim (\nu/\tau_R)^{1/2} = \psi^{1/2} c_s$ be greater than the
upstream velocity, here $v^1 = u^1/u^0$ in the left state. Thus, we require
$v_{\rm char} > v^1 \implies \psi > (v^1/c_s)^2$. For our chosen set of
parameters we have $v_{\rm char} \approx 0.756 > v^1 \approx 0.707$,
and hence we are able to resolve the shock structure. We find that the major
contribution to the shock structure comes from the pressure anisotropy; the role
of the heat conduction inside the shock is marginal. The EMHD theory has
hyperbolic dissipation, where $q$ and $\Delta P$ relax to values
$q_0 \propto \nabla_\mu T, \Delta P_0 \propto \nabla_\mu u_\nu$ over a time
scale $\tau_R$. This leads to structure of length $\sim v^1
\tau_R$ over which the dissipation builds up (figure~\ref{fig:bvp_vs_grim}), and
then reaches the relaxed values $q_0, \Delta P_0$. The theory has higher order
corrections $\sim q u^\mu \nabla_\mu(\tau_R/(\chi P^2)), \Delta P u^\mu
\nabla_\mu(\tau_R/(\rho \nu P))$ that we expect to contribute in strong
nonlinear regimes, and indeed we see that the shock structure differs as we turn
on, and turn off, these terms (fig.~\ref{fig:HO_vs_NO_HO}). However, from
fig.~\ref{fig:HO_vs_NO_HO}, we see that the differences are small. Still, their
presence is required to enforce the second law of thermodynamics.

There is an upper limit to the strength of the shock that can be solved for using the EMHD model. Higher mach number shocks require a larger viscosity (or $\Delta P$) to smoothly connect the left and right states. However, the non-dimensional parameter $\psi$ cannot be arbitrarily large because of an upper bound on the associated characteristic speed $v_{\rm char} \sim \psi^{1/2} c_s < c \implies \psi < (c/c_s)^2$. Beyond this critical value, the theory loses hyperbolicity, and eventually causality and stability. The root of this problem lies in the fact that ultimately, the theory is a second order perturbation $\sim q^2, \Delta P^2$, about an equilibrium and, as the dissipative effects become stronger, the validity of the expansion breaks down.

What happens if we do not resolve the shock? In astrophysical applications, this is almost always the case since there is a large separation between the MHD, and the kinetic spatio-temporal scales. The pressure anisotropy $\Delta P$ is limited to the values allowed by the saturation of kinetic instabilites such as mirror and firehose. For example $\Delta P \sim b^2$, where $b^2$ is the magnetic pressure. This viscosity may not be sufficient to resolve a shock. However, since {\tt grim} is a conservative code, even when shocks are not resolved, the obtained solution asymptotes to the value given by the ideal fluid Rankine-Hugoniot jump conditions a few mean free paths away from the shock.

\subsubsection{Anisotropic Conduction Test}

The EMHD model constrains heat to flow only along the magnetic field lines $q_0 \propto \hat{b}^\mu(\nabla_\mu T + T a_\mu)$. To test this, we set up a temperature perturbation in pressure equilibrium, in Minkowski space-time with sinusoidal background magnetic field lines. The domain is a square box of size $[0, 1] \times [0, 1]$ with periodic boundary conditions. The initial conditions are
\begin{align}
\rho & = 1 - A e^{-r^2/R^2} \\
u    & = 1 \\
u^1  & = u^2 = u^3 = 0 \\
B^1  & = B_0 \\
B^2  & = B_0\sin(2\pi k x^1)
\end{align}
where the amplitude of the perturbation $A = 0.2$, the radius $R = \sqrt{.005}$,
the mean magnetic field $B_0 = 10^{-4}$, and the wavenumber of the magnetic
field $k = 4$. The adiabatic index is set to $\gamma = 4/3$, the relaxation time
scale in the EMHD model $\tau_R = 0.1$, and the thermal diffusivity $\chi = 0.01$.

\begin{figure}[!htbp]
\begin{center}
\includegraphics[width=180mm]{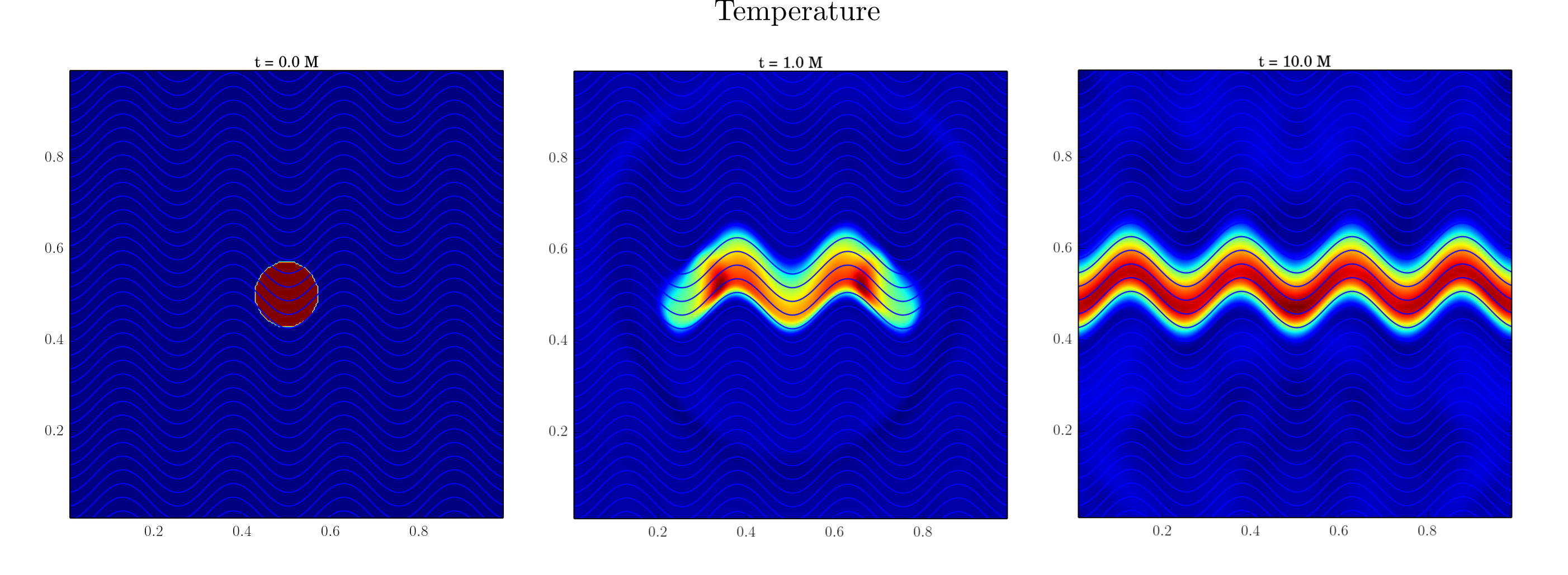}
\end{center}
\caption{Evolution of a temperature perturbation, initially in pressure equilibrium, over sinusoidal magnetic field lines. This test provides a nice visualization of the anistropic transport of the EMHD theory.}
\label{fig:snake_test}
\end{figure}

Since the initial conditions are in pressure equilibrium, they are an exact time independent solution of the ideal MHD equations. However, the EMHD model is sensitive to temperature gradients along field lines, and hence the system should evolve to a state where the plasma becomes isothermal along field lines. This outcome is shown in fig.~\ref{fig:snake_test}, along with the transient state. As the heat flows, it excites sound waves that traverse the domain, eventually reaching the steady solution shown in the last panel in fig.~\ref{fig:snake_test}.

\subsubsection{Firehose Instability}
\begin{figure}[!htbp]
\begin{center}
\includegraphics[width=180mm]{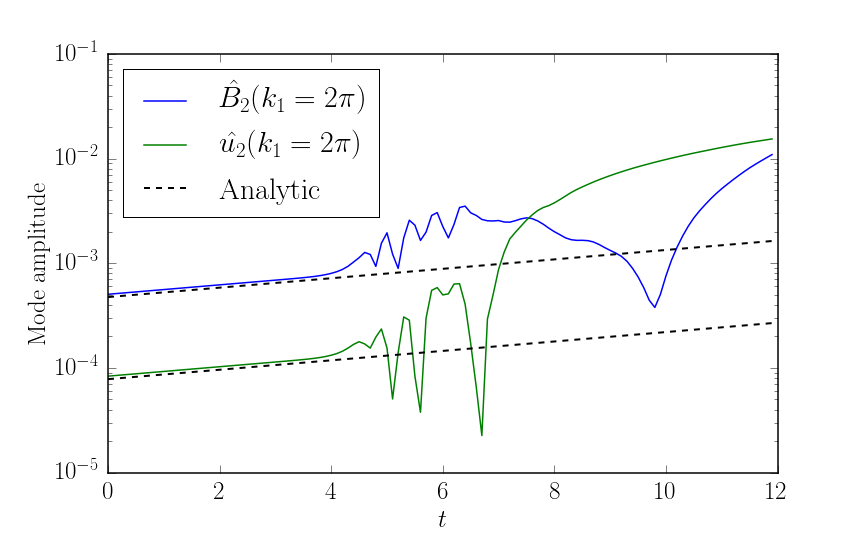}
\end{center}
\caption{Mode growth in the firehose instability test.}
\label{fig:firehose}
\end{figure}

The EMHD model, like Braginskii's theory of weakly collisional anisotropic plasmas, is susceptible to the firehose instability.  If $\Delta P < -b^2$ Alfv\'en waves become unstable and grow at a rate proportional to their wavenumber (see~\cite{chandra2015} for the EMHD result).
To test the linear growth of a firehose-unstable mode, we consider the following initial conditions on a Minkowski background:
\begin{align}
\rho & = 1 \\
u    & = 2 \\
u^1  & = u^3 = 0 \\
u^2 &= A \sin{(2\pi x^1)}\\
B^1  & = 0.1 \\
B^2 & = B \cos{(2\pi x^2)}\\
B^3 & = 0 \\
\Delta P &= -0.011
\end{align}
with $A=0.1628\alpha$ and $B=0.9867\alpha$ chosen so that the perturbation of amplitude $\alpha$ is one of the linearly unstable Alfven modes,
with exponential growth rate $\Gamma=0.1036$. We artificially impose a very slow damping rate $\tau_R=10^6$ to the pressure anisotropy, to avoid
rapid damping of the imposed pressure anisotropy $\Delta P$ towards its equilibrium value $\Delta P \approx 0$ (as the background flow has no shear).

In Fig.~\ref{fig:firehose} we show the evolution of the unstable mode amplitude.  We observe two separate regimes of evolution. First, the unstable mode grows exponentially
at the predicted rate $\Gamma$, in agreement with the linear theory. At later times truncation error seeds perturbations on smaller length scale, which have a
much faster growth rate. Around $t=4$ the growth of the perturbation is dominated by grid-scale modes, which grow much faster than the mode we inserted in the initial conditions, and quickly become nonlinear.

In kinetic theory, the pressure anisotropy saturates at $\Delta P \approx - b^2$.
In astrophysical simulations, we similarly impose a saturation of $\Delta P$ by smoothly reducing $\tau_R$ if  $\Delta P < -b^2$.

\subsubsection{Hydrostatic Conducting Atmosphere}
\begin{figure}[!htbp]
\begin{center}
\includegraphics[width=140mm]{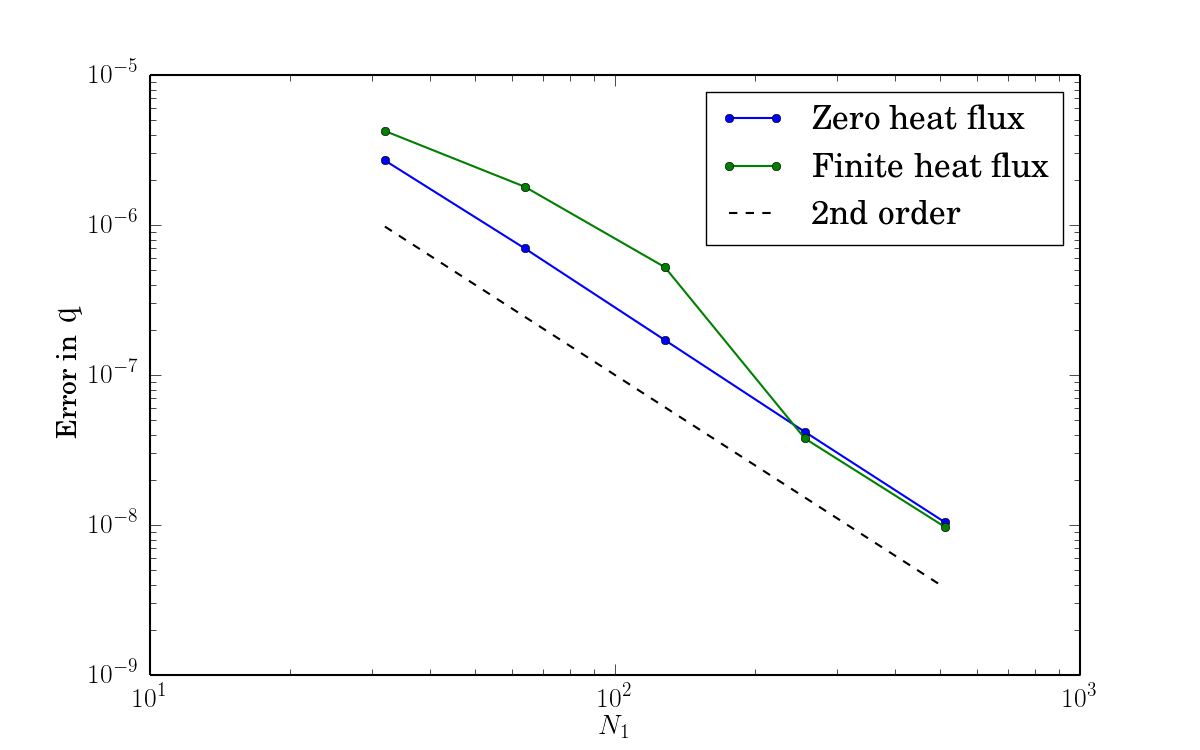}
\end{center}
\caption{Convergence for a hydrostatic atmosphere with zero, and finite heat flux at $t = 10$ $GM/c^3$.}
\label{fig:atmosphere}
\end{figure}

Heat conduction in curved space-times contains qualitatively new
features when compared to Minkowski space-time because the heat flux is
driven by \emph{red-shifted} temperature gradients $q_0 \propto \hat{b}^\mu(\nabla_\mu \Theta + \Theta a_\mu)$ where $a^\lambda = u^\nu \nabla_\nu u^\lambda \equiv u^\nu\partial_\nu u^\lambda + \Gamma^\lambda_{\mu \nu} u^\mu u^\nu$ is the four-acceleration. For a fluid at rest in a stationary spacetime this simplifies to $q_0 \propto \partial_i(\Theta \sqrt{-g_{00}})/\sqrt{-g_{00}}$. Thus, a zero heat flux configuration corresponds to $\partial_i(\Theta \sqrt{-g_{00}}) = 0$, and not $\partial_i\Theta = 0$. A fluid element deep in a gravitational potential well requires greater internal energy in order to stay in thermal equilibrium with a fluid element outside the potential well. We test this effect with a hydrostatic fluid configuration in a Schwarzschild metric in the domain $(R, \theta) = [200\;M, 300\;M] \times (0, \pi/2)$. The equations of hydrostatic equilibrium are
\begin{align}
\frac{\partial P}{\partial x^1} & = -(\rho + u + P)\frac{\partial \ln \sqrt{-g_{00}}}{\partial x^1} \label{eq:atmosphere_pressure}\\
\frac{\partial (q\sqrt{g*g_{00}})}{\partial x^1} & = \sqrt{-g} T^\kappa_\lambda \Gamma^\lambda_{\nu \kappa} \label{eq:atmosphere_heat_flux} \\
\frac{\partial(\Theta \sqrt{-g_{00}})}{\partial x^1} & = q \label{eq:atmosphere_temperature}
\end{align}
where (\ref{eq:atmosphere_pressure}) is the momentum conservation equation in
the radial direction, (\ref{eq:atmosphere_temperature}) is the energy equation,
and (\ref{eq:atmosphere_heat_flux}) is the evolution equation for the heat flux
(\ref{eq:qevol}), simplified in the presence of a radial magnetic field, and the
absence of a radial velocity ($u^r = 0$) . The above equations are
one-dimensional ODEs in the radial direction which we integrate outwards between
two concentric spheres, starting with $(P_0, \Theta_0, q_0)$ at the inner
boundary. The above equations are augmented by the ideal gas equation of state
$u = P_g/(\gamma - 1)$, with $\gamma = 4/3$, and $\rho = P/\Theta$ to determine $u$ and $\rho$ respectively. The resulting (semi-)analytic solutions are then used as initial conditions in {\tt grim}. If the numerical implementation is correct, {\tt grim} should maintain the equilibrium.  We consider two cases, (1) $q_0 = 0 \implies q = 0$, which is a system in thermal equilibrium, and (2) $q_0 \neq 0 \implies q \neq 0$, corresponding to a system that is conducting heat radially outwards.  Fig.~\ref{fig:atmosphere} shows the errors at the final time of the evolution falling off at the expected order for both cases.

\subsubsection{Bondi Inflow}

Spherical accretion onto a non-spinning black hole is a common test of general relativistic hydrodynamics code. It is a rare case of a non-trivial configuration for
which a steady-state solution can be obtained analytically. For this test, we use as background flow the well-known solution for a spherical accretion flow
around a non-spinning black hole of mass $M=1$ due to \citealt{Michel}. This solution has a sonic point which we places at $r_s = 8 GM/c^2$. We also
add a radial magnetic field $B^r = 1/\sqrt{-g}$, which does not modify the hydrodynamics equilibrium.

The Bondi inflow solution has a non-trivial $u^r$. The presence of a finite inflow velocity exercises all the time-independent terms in the EMHD equations for $q$ and $\Delta P$, including higher order terms that are identically zero in a hydrostatic solution. We obtain reference solutions by ignoring backreaction of the dissipation onto the fluid flow and integrate one-dimensional ODEs in the radial direction for $q$ and $\Delta P$. We then use these solutions to check {\tt grim} results obtained with backreaction turned off.

\begin{figure}[!htbp]
\begin{center}
\includegraphics[width=140mm]{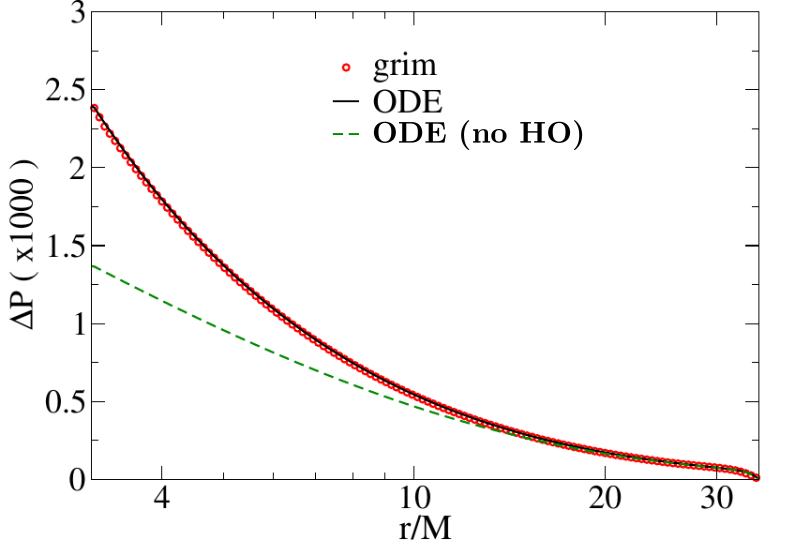}
\end{center}
\caption{Pressure anisotropy at $t=1000 GM/c^3$ for a {\tt grim} evolution of spherical accretion in the EMHD model, without backreaction of the pressure anisotropy
onto the flow (red circles). A numerical integration of the analytical solution
  is shown as a solid black line. The analytical solution without higher order (HO)
terms (i.e. simply damping the advected $\Delta P$ to $\Delta P_0$ on a timescale $\tau_R$) is shown as a dashed green line.}
\label{fig:BondiConduction}
\end{figure}

Fig.~\ref{fig:BondiConduction} shows the value of the pressure anisotropy obtained at time $t=1000GM/c^3$ of a {\tt grim} evolution,
and from the simpler ODE integration on top of the steady-state fluid background. The two are in very good agreement.
We also note that spherical accretion is an interesting case in which the high-order terms in the evolution of $q$ and $\Delta P$ and choice of
damping timescale $\tau_R$ do change the flow by order unity. Indeed, the rescaled pressure anisotropy $\Delta \tilde P$ is damped towards its relaxed value in Braginskii's theory
($\Delta \tilde P_0$), as well being advected with the flow. In this test problem, the pressure anisotropy varies rapidly with radius, but the radial velocity of the flow
is also large. For $\tau_R \gtrsim r/u^r$, the pressure anisotropy can thus remain significantly smaller than its Braginskii target. Fig.~\ref{fig:BondiConduction} uses $\tau_R=30$
everywhere.

\subsection{Ideal MHD Tests}

EMHD reduces to ideal MHD in the limit of vanishing diffusion coefficients ($\chi, \nu \rightarrow 0$), resulting in zero dissipation ($q, \Delta P \rightarrow 0)
$\footnote{The limits $\chi, \nu \rightarrow 0$ need to be taken carefully because diffusion coefficients appear in the denominator of the higher order terms ($\sim \tilde{q} \nabla_\mu u^\mu, \Delta \tilde{P} \nabla_\mu u^\mu$) in (\ref{eqn:q_evol_eqn_rescaled}), and (\ref{eqn:dP_evol_eqn_rescaled}), where $\tilde{q} \sim q/\sqrt{\chi}, \Delta \tilde{P} \sim \Delta P/\sqrt{\nu}$ .  To obtain the correct limit, rescale (\ref{eqn:q_evol_eqn_rescaled}) by $\sqrt{\chi}$, and then take $\chi \rightarrow 0$, leading to $q \rightarrow 0$. The limit $\Delta P \rightarrow 0$ follows similarly.}
Therefore, any code that solves the EMHD equations should also be able to handle ideal MHD. To check this, we subject {\tt grim} to ideal MHD shock tests in order to check its shock capturing ability. To solve the ideal MHD equations, we simply ignore the evolution of the heat flux (\ref{eqn:q_evol_eqn_rescaled}), and the pressure anisotropy (\ref{eqn:dP_evol_eqn_rescaled}), as well as the relevant terms in the stress-energy tensor (\ref{eq:fullSEtens}). This leads to the assembly, and inversion of a $5 \times 5$ Jacobian (for the variables $\{\rho, u, u^1, u^2, u^3\}$) in the residual-based root finder, as opposed to a $7 \times 7$ Jacobian for EMHD.

We have successful tested {\tt grim} on the following ideal MHD problems 1) \cite{komissarov1999} shock tests, 2) relativistic Orzag-Tang \citep{beckwith2011}, 3) diagonal transport of an overdensity \citep{harm}, 4) low, and medium magnetized cylindrical blast wave \citep{komissarov1999}, 6) steady-state hydrodynamic torus \citep{fishbone}. The Riemann solver in {\tt grim} is identical to that used in {\tt harm}, therefore we are prone to all of the known issues of the {\tt harm} scheme. Specifically, the Local Lax Friedrichs (LLF) flux that we use leads to excess diffusion at contact discontinuities when compared to schemes that explicitly model the discontinuity, like HLLC \citep{HLLC}.

\subsubsection{Komissarov shock tests}
\begin{figure}[!htbp]
\begin{center}
\includegraphics[width=180mm]{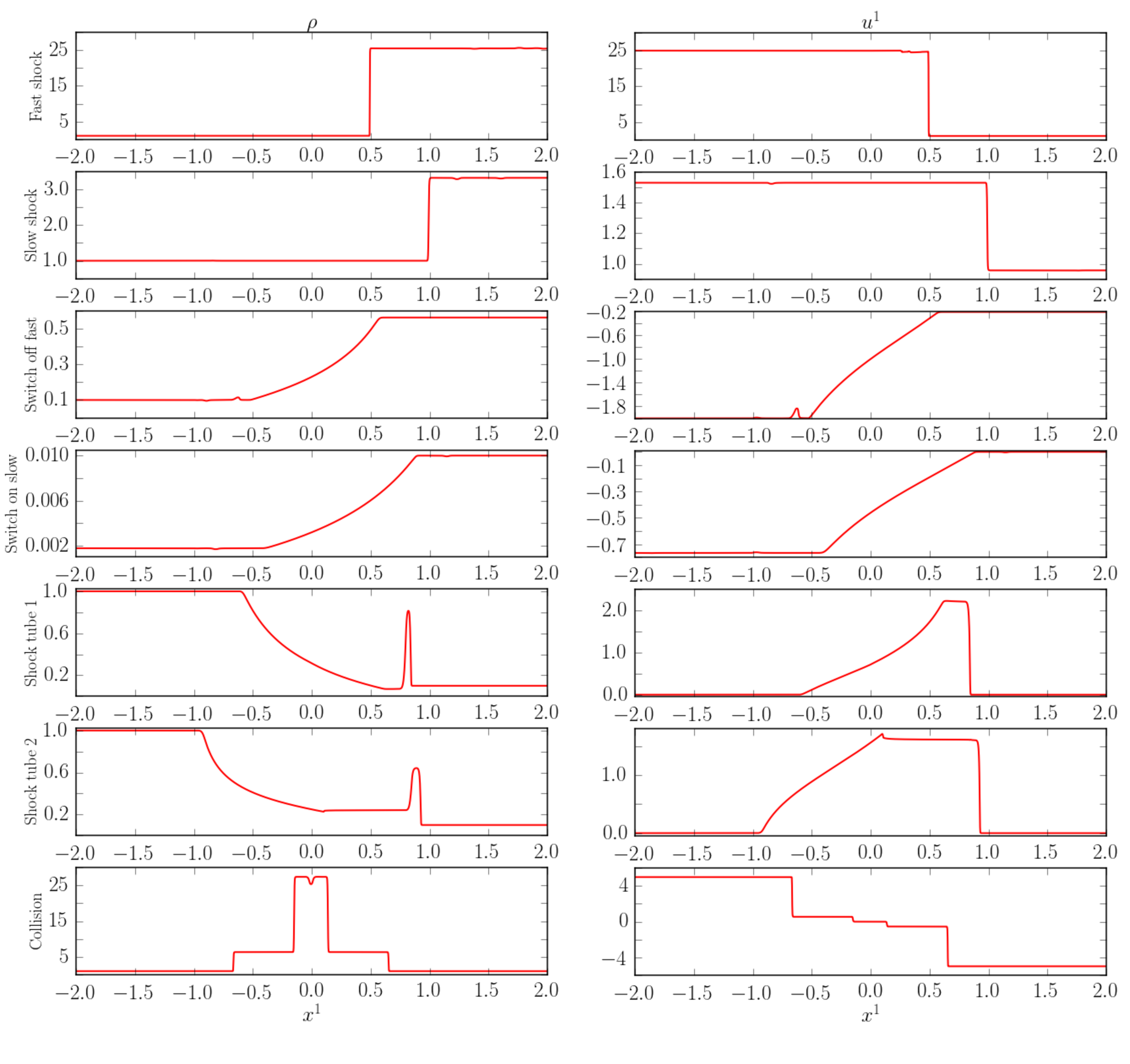}
\end{center}
\caption{Density $\rho$ (left panel), and four-velocity $u^1$ (right panel) for each of the Komissarov shock tests.}
\label{fig:komissarov}
\end{figure}

\cite{komissarov1999} formulated a series of one-dimensional nonlinear MHD solutions that are designed to check a codes ability to correctly handle shocks and rarefactions. We ran the following cases: (1) fast shock, (2) slow shock, (3) switch-off fast, (4) switch-on slow, (5) shock-tube 1, (6) shock tube 2, and (7) collision. We ran each case with 2048 grid zones in a domain $[-2, 2]$ with a minmod limiter (which in our implementation is the generalized minmod limiter, with slope set to one), and a courant factor of 0.2. As is shown in Fig.~\ref{fig:komissarov}, we correctly reproduce the expected results. 

\section{Applications} \label{section:applications}
We describe three example applications that highlight the new physics in the
EMHD model: (1) Buoyancy instabilities in weakly collisional plasmas and (2)
radiatively inefficient accretion flows around supermassive black holes. We
study these in global 3D domains using coordinates by \cite{MADJet,
sasha2016} that smoothly cylindrify the grid zones near the poles. This
mollifies the severe time step constraints in the azimuthal ($\phi$) direction.

\subsection{Buoyancy Instabilities}

Weakly collisional plasmas are subject to instabilities not present in ideal plasmas, due to the presence of anisotropic dissipation. An ideal plasma that satisfies the Schwarzschild criterion $ds/dz > 0$ is convectively stable. However, this is not the case when the heat flux is constrained to be parallel to magnetic field lines.

\begin{itemize}
\item When the temperature \emph{decreases} outwards $dT/dz < 0$ against gravity in the presence of magnetic field lines that are \emph{perpendicular} to the temperature gradient \footnote{When the field lines are aligned along the negative temperature gradient $dT/dz < 0$, the system is MTI stable.}, the plasma is unstable to the \emph{Magneto-Thermal Instability} (MTI) \citep{Balbus2000}.

\item When the temperature \emph{increases} outwards $dT/dz > 0$ against gravity in the presence of magnetic field lines that are \emph{parallel} to the temperature gradient\footnote{When the field lines are aligned perpendicular to the positive temperature gradient $dT/dz > 0$, the system is HBI stable.}, the plasma is unstable to the \emph{Heat flux driven Buoyancy Instability} (HBI) \citep{Quataert2008}.
\end{itemize}

Both instabilities require weak magnetic fields, else they are suppressed by
strong magnetic tension. The linear growth and nonlinear saturation of these
instabilities have been studied in-depth in the non-relativistic regime using
the \citealt{Braginskii1965} model for weakly collisional plasmas. The EMHD
model reduces to the Braginskii model in the non-relativistic limit when $\tau_R \rightarrow 0$. We expect to see the MTI and the HBI in the EMHD model, and indeed we do.  Below we describe the setups and the linear and the nonlinear regimes for both instabilities.

We use hydrostatic Schwarzschild-stable initial conditions in a Schwarzschild
metric. We want to be able to control the sign of the temperature gradient, so
that the system is either MTI or HBI unstable. To do so, we set the initial $P_g
= K \rho^\Gamma$, where $K$ is a constant, and $\Gamma$ is the polytropic index.
$P$ is solved for using hydrostatic equilibrium (\ref{eq:atmosphere_pressure}),
which then yields $\rho = (P_g/K)^{1/\Gamma}$ and $u$, using $u = P_g/(\gamma - 1)$.

A Schwarszchild stable equilibrium requires $ds/dr > 0 \implies \Gamma < \gamma$. $\Gamma$ can be changed to obtain either a positive temperature gradient $dT/dr > 0$ ($\Gamma < 1$), or a negative temperature gradient $dT/dr < 0$  ($\Gamma > 1$). For MTI, we set $K = 10^{-4}$ and $\Gamma = 4/3$, while for HBI we set $K = 0.05$ and $\Gamma = 1/2$.

We set $\chi = c_s R$, where $R$ is radius and $c_s = \sqrt{\gamma P/\rho}$, as
in \citealt{sharma2008}. The EMHD model has an additional parameter, the
relaxation time scale, set via $\tau_R = R/c_s$.

\subsubsection{Magneto-Thermal Instability}

The MTI requires magnetic field lines perpendicular to the temperature gradient for maximal growth. Therefore, we perform the simulation in a half sphere $(R, \theta, \phi) \in [200M, 300M] \times (0, \pi) \times [0, \pi]$, and initialize with a weak azimuthal magnetic field $B_\phi = 10^{-3}/\sqrt{-g}$. We use Dirichlet boundary conditions in $R$, and $\theta$ for the density $\rho$, pressure $P$, and internal energy $u$. This results in constant temperature boundaries, which continuously drive the instability. We use insulating boundary conditions for the magnetic fields, i.e. set them to zero in the boundaries. The $\phi$ boundaries are periodic for all variables.

\begin{figure}[!htbp]
\begin{center}
\includegraphics[width=180mm]{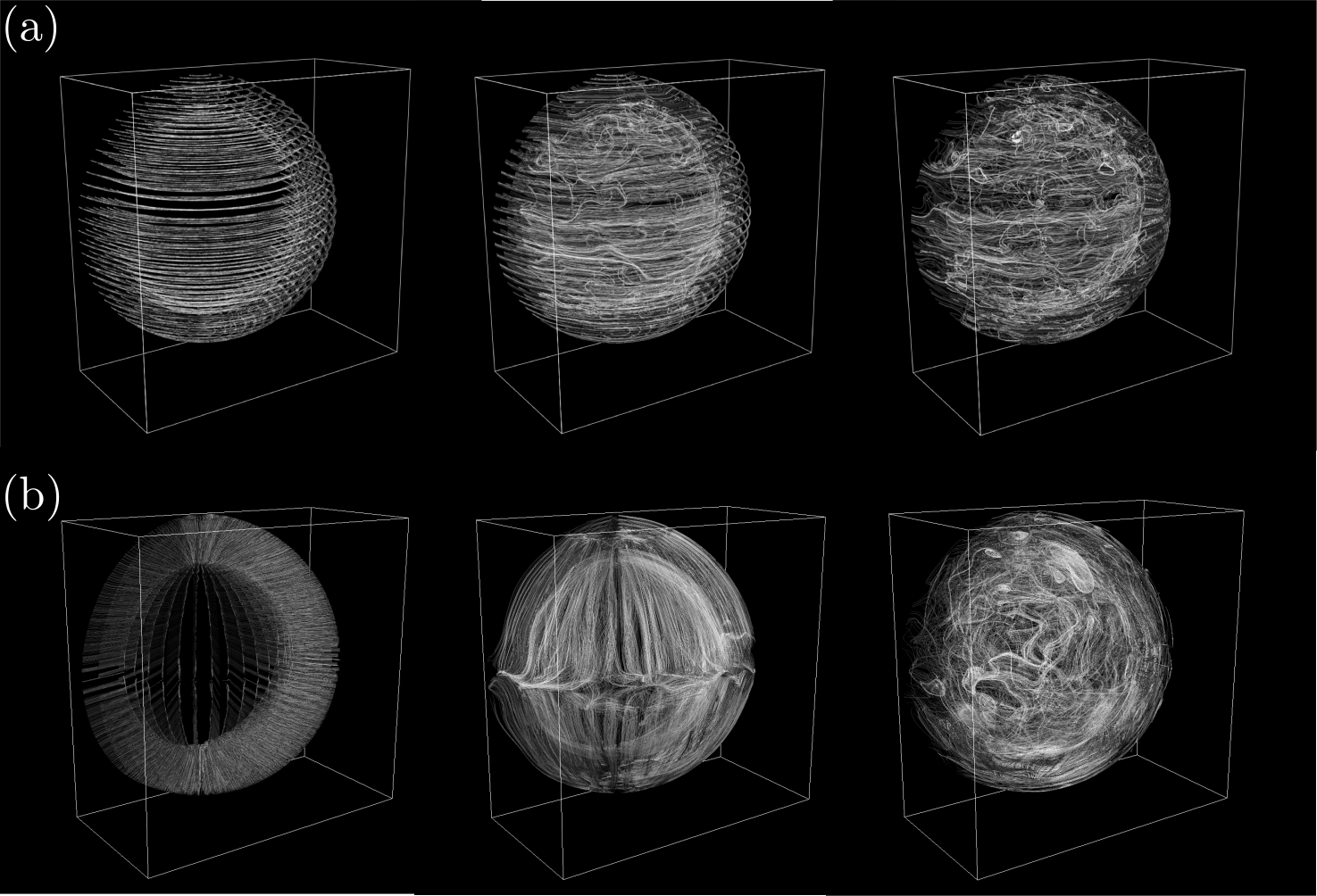}
\end{center}
  \caption{From left to right: Evolutions exhibiting Numerical evolution of (a) the magnetothermal instability (MTI) and (b) the heat flux driven
  buoyancy instability (HBI). Case (a) is initialized with a purely azimuthal field,
  and a temperature profile decreasing outwards, which is unstable to the MTI,
  and leads to an exponential growth in the radial component of the magnetic field.
  Case (b) starts with a purely radial field, and a temperature profile
  increasing outwards, which is unstable to the HBI, and leads to an exponential
  growth in the perpendicular component of the field. The  free parameters of the
  EMHD theory are the same in both cases. Both cases use $128 \times 128 \times 128$ grid zones in $(R, \theta, \phi) \in [200M,
  300M] \times (0, \pi) \times [0, \pi)$.}
\label{fig:MTI}
\end{figure}

\begin{figure}[!htbp]
\begin{center}
\includegraphics[width=150mm]{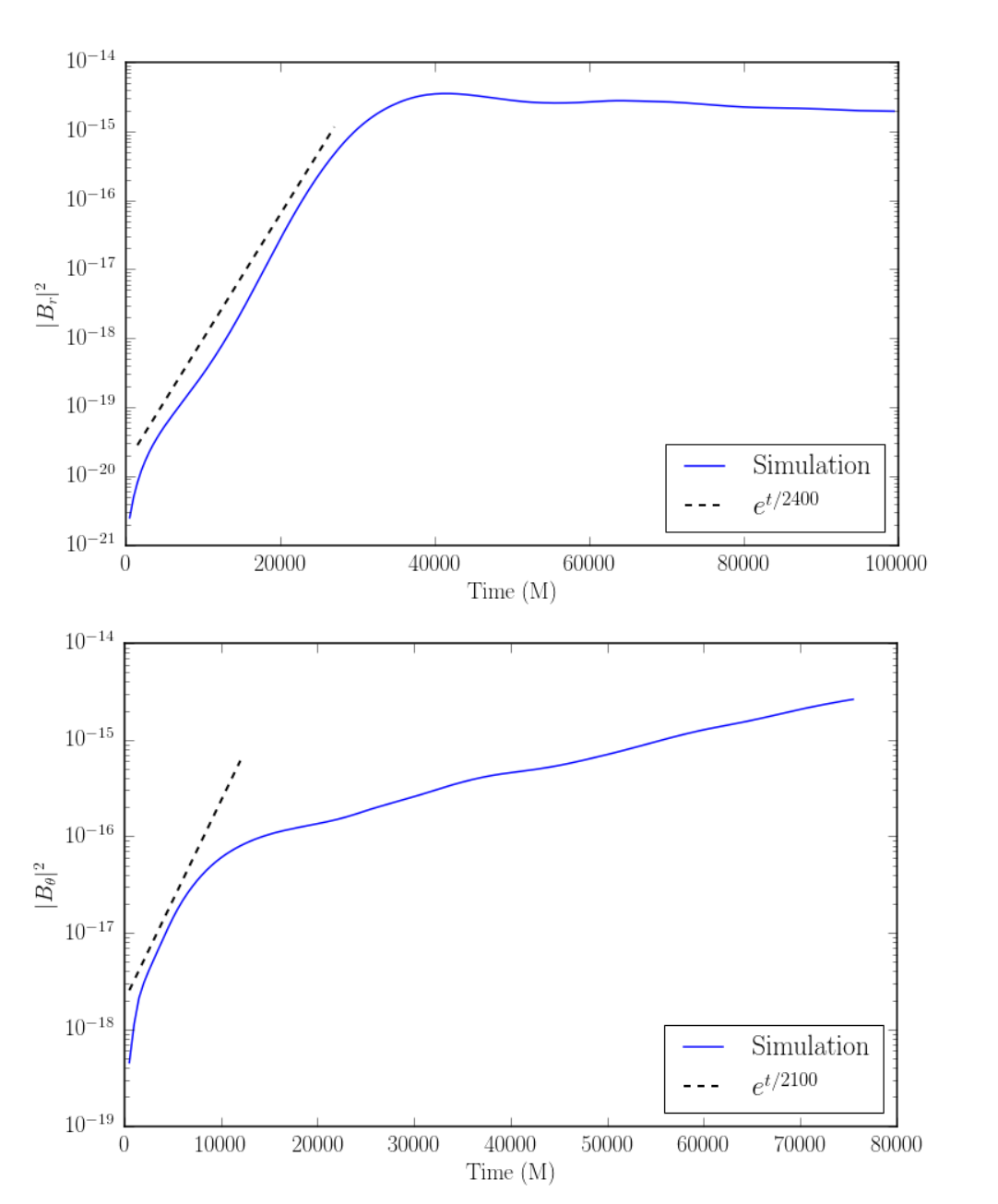}
\end{center}
  \caption{Top (bottom) panel : Growth of the radial ($\theta$) component of the
  magnetic field in the 3D setup to study the MTI (HBI). The dotted line
  corresponds to an exponential growth with time scale $\sim 2400$ M ($2100$ M).
  In the limit where the conduction time scale is the fastest, as in our setups,
  the instabilities grow on a dynamical time scale, which for $R = 200$ M is
  $\sim 2500$ M ($2000$ M).}
\label{fig:MTI_growth}
\end{figure}

\begin{figure}[!htbp]
\begin{center}
\includegraphics[width=150mm]{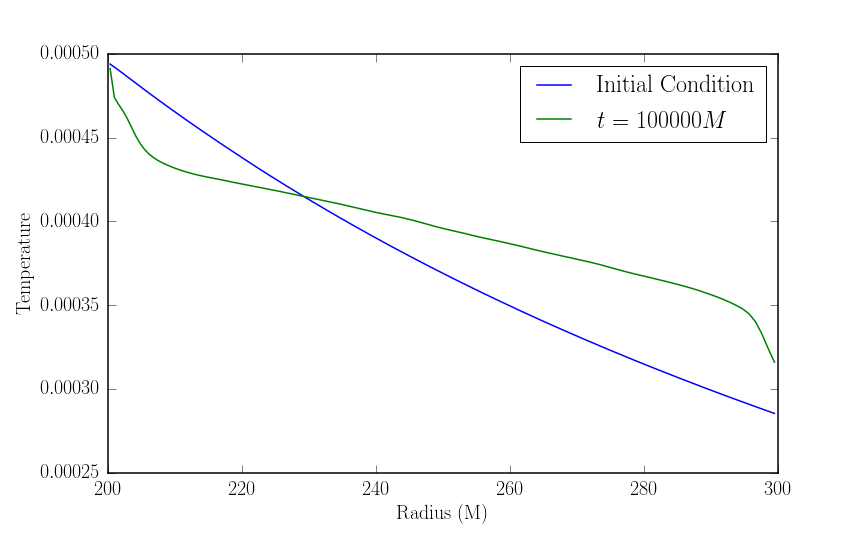}
\end{center}
\caption{Initial and final radial temperature profiles (averaged over $(\theta,
  \phi)$) in the saturated state of the MTI. The instability is driven by the
  boundaries at $R = 200, 300$ M, which are held at fixed temperatures. The
  fixed temperature boundaries resist the flattening of the temperature profile
  due to the MTI, thus creating kinks in the temperature profile close to
  the radial boundaries.}
\label{fig:MTI_temperature_profile}
\end{figure}

The initial conditions have zero heat flux $q = 0$, as well as $q_0 \sim \bh^\mu (\nabla_\mu \Theta + \Theta a_\mu) = 0 $. We seed the simulation with small amplitude, $\sim 4 \%$, fluctuations in $u^1$. These lead to small scale corrugations of the field lines whose radial component is exponentially amplified due to the MTI (\ref{fig:MTI_growth}a). Eventually, there is vigorous convection (fig.~\ref{fig:MTI}a), and a net heat flux between the radial boundaries, leading to a flattened temperature profile in the bulk of the domain (fig.~\ref{fig:MTI_temperature_profile}).  This is consistent with expectations from nonlinear evolution of the nonrelativistic MTI.  

\subsubsection{Heat-Flux Driven Buoyancy Instability}

The HBI requires magnetic field lines to be aligned with the temperature gradient for maximal growth, and so we seed the simulation with radial field lines $B^r = 10^{-3}/\sqrt{-g}$. The spatial domain is the same 3D half-sphere of the MTI setup. 
The boundary conditions, $\chi$, and $\tau_R$ are also identical to the MTI case.

\begin{figure}[!htbp]
\begin{center}
\includegraphics[width=100mm]{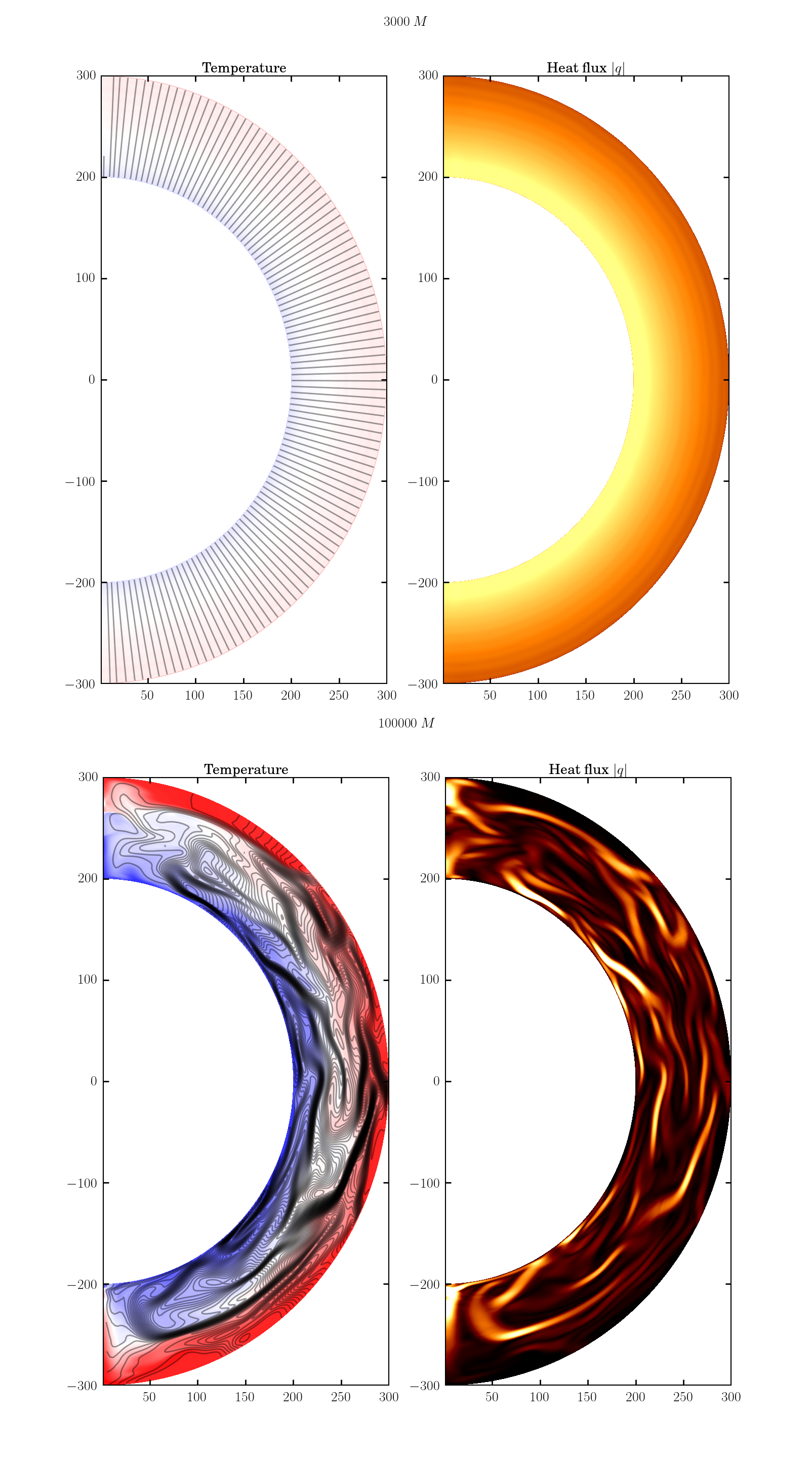}
\end{center}
\caption{Above: The intermediate state of the HBI in a global 2D setup illustrating the
  finite radial heat flux that develops due to the presence of radial field
  lines connecting the constant temperature boundaries at $R = 200$ and $R =
  300$ M. The initial conditions have zero heat flux, and are not shown here.
  Below: The saturated state of the HBI that suppresses the radial heat
  flux of the intermediate state.}
\label{fig:HBI}
\end{figure}

The initial conditions have $q = 0$, but $q_0 \sim \bh^\mu (\nabla_\mu \Theta +
\Theta a_\mu)\neq 0$. The heat flux $q$ relaxes to $q_0$ over a timescale
$\tau_R$, leading to a finite radial heat flux. This heat flux feeds the HBI, which grows by kinking the field lines, and leads to an exponential growth of the radial component of the magnetic field. In the saturated state there is suppression of the heat flux below $q_0$. Fig.~\ref{fig:HBI} shows the intermediate state $q_0$, which is unstable to the HBI, and the final saturated state.

\subsection{Radiatively Inefficient Accretion Flow}

\begin{figure}
\begin{center}
\includegraphics[width=0.45\textwidth]{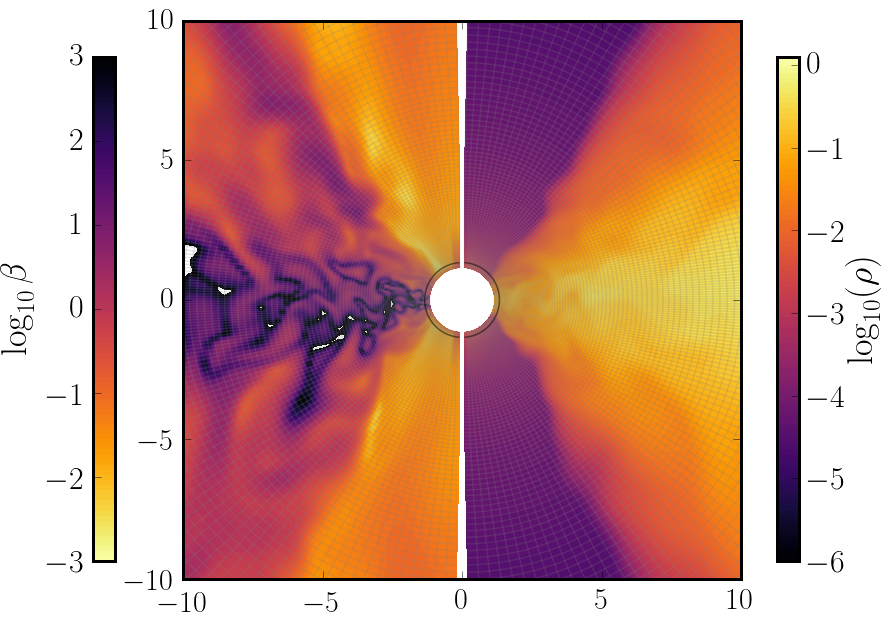}
\includegraphics[width=0.47\textwidth]{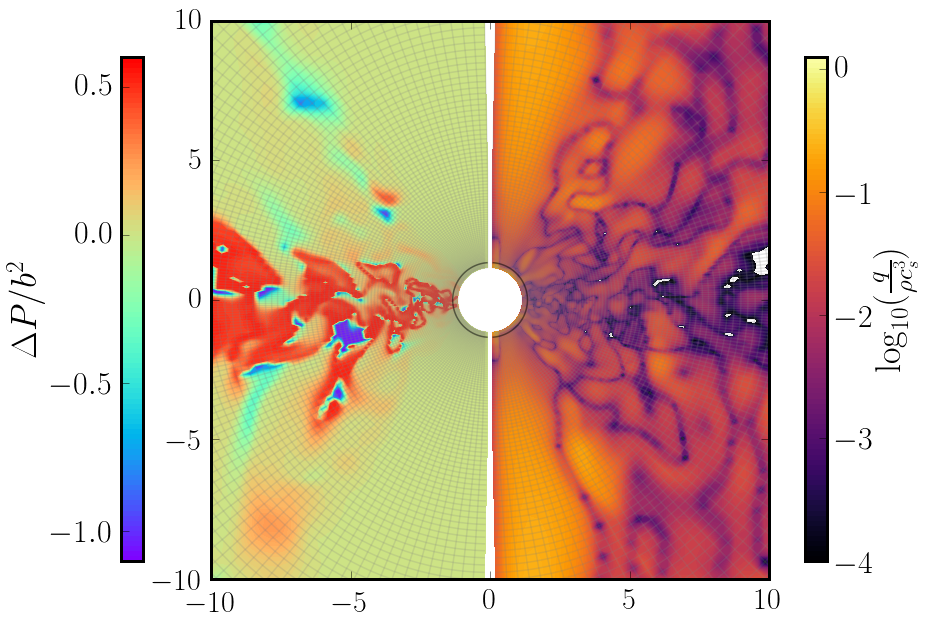}
\end{center}
\caption{Vertical slice in the 3D evolution of a torus in the EMHD model, at time $t=1240GM/c^3$.
We show the plasma parameter $\beta=2P/b^2$, density $\rho$, pressure anisotropy scaled to the magnetic energy $\Delta P/b^2$,
and heat flux scale to the {\it free-streaming} heat flux $q/(\rho c_s^3)$.}
\label{fig:torus}
\end{figure}

The first astrophysical targets for the {\tt grim} code are slowly accreting supermassive black holes.
For a black hole with an accretion rate $\dot{M} \lesssim 0.01 \dot{M}_{\rm Edd}$ ($\dot{M}_{\rm Edd} \equiv$ Eddington
rate), we expect the surrounding accretion disk to be formed of a weakly collisional, magnetized plasma whose evolution is better
approximated by our EMHD model than by the equations of ideal magnetohydrodynamics. We have already
used the {\tt grim} code to study the evolution of an accretion disk in the EMHD model in global, axisymmetric simulations~\cite{}.
The current version of the {\tt grim} code has also been tested on short preliminary evolutions of accretion disks in 3D, at low resolution.

In both cases, we find that the pressure anisotropy in the disk grows to values comparable to the magnetic pressure in the disk,
reaching the mirror instability threshold. The closure used in our EMHD model then forces $\Delta P$ to saturate
at $\approx b^2/2$. Longer, higher-resolution simulations are necessary to fully assess the impact of the EMHD model
on the dynamics and energy budget of the system, and will be performed as sufficient computational resources become available.

Fig.~\ref{fig:torus} shows a snapshot of such a 3D evolution at $t=1240 GM/c^3$. The simulation was started from a hydrodynamical
equilibrium torus \cite{fishbone} around a spinning black hole ($a = 0.9375$), 
seeded with a single loop of poloidal magnetic field. The initial amplitude of the plasma
parameter $\beta \equiv 2P/b^2$ is $\sim 100$ in the inner disk, and $\beta \gtrsim 15$ everywhere. We see growth of magnetic turbulence due to the magnetorotational instability, and growth of the pressure anisotropy to the mirror instability threshold $\Delta P = b^2/2$. The heat
flux is $\sim 10\%$ of its free-streaming value, a much larger effect than in earlier axisymmetric simulations \citep{foucart2016}.

\section{Conclusion} \label{section:conclusion}

Low luminosity black hole accretion flows ($L \ll L_{\rm edd}$) are expected to
be collisionless, so anisotropic dissipative effects can be important.
Understanding the disk structure, and predicting observables requires the
nonlinear solutions of relativistic dissipative theories in strongly curved
space-times. Numerical codes so far can only evolve perfect fluids, with no heat
conduction or viscosity. The algorithms developed for perfect fluids do not work
for relativistic dissipative theories, because dissipation in the relativistic
case is sourced by \emph{spatio-temporal} gradients of the thermodynamic
variables, as opposed to just spatial gradients in the non-relativistic case. In
this paper, we have formulated and implemented a new scheme that can handle this
situation and is physics-agnostic. We implement the scheme in a new code {\tt
grim}, which we then use to integrate the EMHD theory of anisotropic
relativistic dissipation. The numerical solutions obtained have been checked
against various analytic and semi-analytic solutions of the EMHD theory in both
Minkowski and Schwarszchild spacetimes, in linear as well as in non-linear
regimes.

The algorithm is the same as in \citealt{foucart2016} that has been used to
study axisymmetric radiatively inefficient accretion flows, although here the
code has been generalized to work in 3D, and now has the ability to run on
either CPUs or GPUs.  Thus we are able to make full use of the various node
architectures in current and future generations of supercomputers. We use a
performance model to show that the implementation is near-optimal, with the code
achieving a significant fraction ($\sim 70-80 \%$) of peak machine bandwidth. This,
we show is crucial, because the performance of nonlinear solver that is at the
heart of {\tt grim} is primarily dependent on the machine bandwidth.

As example applications we have studied the magneto-thermal instability (MTI)
and the heat flux driven buoyancy instability (HBI) in global 3D domains with a
Schwarzschild metric, and evolved them to a nonlinear saturated state. Finally,
we performed preliminary EMHD evolutions of a hydrodynamically stable torus in
3D, around a spinning (Kerr) black hole.

\acknowledgements
We thank Ben Ryan, Sasha Tchekhovskoy, Sean Ressler, Eliot Quataert, and Jim Stone for discussions as well as all the members of the 
horizon collaboration for their advice and encouragement (see {\tt
horizon.astro.illinois.edu}).  The horizon collaboration is supported in part by
NSF. We thank Pavan Yalamanchili at {\tt ArrayFire} for his help with
optimization. Support for this work was provided by NASA through an Einstein
Postdoctoral Fellowship grant numbered PF4-150122 awarded to FF by the Chandra
X-ray Center, which is operated by the Smithsonian Astrophysical Observatory for
NASA under contract NAS8-03060. MC was supported by an Illinois Distinguished
Fellowship from the University of Illinois and by NSF grant AST-1333612. MC
thanks Eliot Quataert for a Visiting Scholar appointment at the University of
California, Berkeley, where part of this work was done.  CFG was supported by NSF grant AST-1333612, a Simons Fellowship, and a visiting fellowship at All Souls College, Oxford.  CFG is also grateful to Oxford Astrophysics for a Visiting Professorship appointment.
This work was made possible by computing time granted by UCB on the Savio cluster.

\bibliographystyle{apj}
\bibliography{local}

\end{document}